\documentclass[]{aa}  

\usepackage{graphicx}
\bibpunct{(}{)}{;}{a}{}{,}
\usepackage{txfonts}
\usepackage{amssymb}
\usepackage{multirow}
\usepackage{float}
\usepackage{gensymb}
\usepackage{color}
\usepackage{rotating}
\usepackage{hyperref}
\hypersetup{backref=true, pagebackref=true, hyperindex=true, breaklinks=true,colorlinks=true,urlcolor=blue, linkcolor=blue, citecolor=blue,pagecolor=red, bookmarks=true, bookmarksopen=true}

\begin{document}

   \title{Evolution of protoplanetary disks from their taxonomy\\in scattered light: Group I vs Group II \thanks{Based on observations collected at the European Organisation for Astronomical Research in the Southern Hemisphere, Chile, under program number 095.C-0658(A)}}


   \author{A.\,Garufi \inst{\ref{Cantoblanco},\ref{ETH}}
   \and G.\,Meeus\inst{\ref{Cantoblanco}}
   \and M.\,Benisty\inst{\ref{IPAG}}
   \and S.P.\,Quanz\inst{\ref{ETH}}
   \and A.\,Banzatti\inst{\ref{Tucson}, \ref{SpaceTelescope}}
   \and M.\,Kama\inst{\ref{Cambridge}}
   \and H.\,Canovas\inst{\ref{Cantoblanco}}
   \and C.\,Eiroa\inst{\ref{Cantoblanco}}
   \and H.M.\,Schmid\inst{\ref{ETH}}
   \and T.\,Stolker\inst{\ref{AMSTERDAM}}
   \and A.\,Pohl\inst{\ref{MPIA}}
   \and E.\,Rigliaco\inst{\ref{Padova}}
   \and F.\,M\'{e}nard\inst{\ref{IPAG}}
   \and M.\,R.\,Meyer\inst{\ref{Michigan}, \ref{ETH}}
   \and R.\,van Boekel\inst{\ref{MPIA}}
   \and C.\,Dominik\inst{\ref{AMSTERDAM}}
         }

 \institute{Universidad Auton\'{o}noma de Madrid, Dpto. F\'{i}sica Te\'{o}rica, M\'{o}dulo 15, Facultad de Ciencias, Campus de Cantoblanco, E-28049 Madrid, Spain. \label{Cantoblanco}
              \email{antonio.garufi@inv.uam.es}
 \and Institute for Astronomy, ETH Zurich, Wolfgang-Pauli-Strasse 27, CH-8093 Zurich, Switzerland \label{ETH} 
   \and Univ. Grenoble Alpes, Institut de Plan\'{e}tologie et d'Astrophysique de Grenoble (IPAG, UMR 5274), F-38000 Grenoble, France \label{IPAG}
   \and Lunar and Planetary Laboratory, The University of Arizona, Tucson, AZ 85721, USA \label{Tucson}
  \and Space Telescope Science Institute, 3700 San Martin Drive, Baltimore, MD 21218, USA \label{SpaceTelescope}
  \and Institute of Astronomy, Madingley Rd, Cambridge, CB3 0HA, UK \label{Cambridge}
   \and Astronomical Institute Anton Pannekoek, University of Amsterdam, PO Box 94249, 1090 GE Amsterdam, The Netherlands \label{AMSTERDAM}  
   \and Max Planck Institute for Astronomy, K\"{o}nigstuhl 17, 69117 Heidelberg, Germany \label{MPIA}
    \and INAF - Osservatorio Astronomico di Padova, Vicolo dell'Osservatorio 5, 35122 Padova, Italy \label{Padova}    
   \and University of Michigan, Department of Astronomy, 1085 S.\ University, Ann Arbor, MI 48109 \label{Michigan}
             }

   \date{Received - / Accepted -}

 
  \abstract
   {High-resolution imaging reveals a large morphological variety of protoplanetary disks. To date, no constraints on their global evolution have been found from this census. An evolutionary classification of disks was proposed based on their IR spectral energy distribution, with the Group I sources showing a prominent cold component ascribed to an earlier stage of evolution than Group II.    }
   {Disk evolution can be constrained from the comparison of disks with different properties. A first attempt at disk taxonomy is now possible thanks to the increasing number of high-resolution images of Herbig Ae/Be stars becoming available.} 
   {Near-IR images of six Group II disks in scattered light were obtained with VLT/NACO in Polarimetric Differential Imaging, which is the most efficient technique for imaging the light scattered by the disk material close to the stars. We compare the stellar/disk properties of this sample with those of well-studied Group I sources available from the literature.}
   {Three Group II disks are detected. The brightness distribution in the disk of HD163296 indicates the presence of a persistent ring-like structure with a possible connection with the CO snowline. A rather compact ($<100$ AU) disk is detected around HD142666 and AK Sco. A taxonomic analysis of 17 Herbig Ae/Be sources reveals that the difference between Group I and Group II is due to the presence or absence of a large disk cavity ($\gtrsim$ 5 AU). There is no evidence supporting the evolution from Group I to Group II. }
   {Group II disks are {not} evolved versions of the Group I disks. Within the Group II disks, very different geometries exist (both self-shadowed and compact). HD163296 could be the primordial version of a typical Group I disk. Other Group II disks, like AK Sco and HD142666, could be smaller counterparts of Group I unable to open cavities as large as those of Group I.}

\keywords{stars: pre-main sequence --
                planetary systems: protoplanetary disks --
                ISM: individual object: HD144432, HD163296, HD152404, AK Sco, HD144668, HD142666, HD150193, HD145263, HD34282, HD135344B, SAO206462, HD169142, HD97048, HD100453, HD142527, HD139614, HD36112, MWC758, HD100546, HD179218   -- 
                Techniques: polarimetric
               }

\authorrunning{Garufi et al.\,2017}

\titlerunning{The evolution of protoplanetary disks, Group I vs Group II}

   \maketitle


\section{Introduction} \label{Sect_Introduction}
Recent space missions and high-contrast imagers have revealed a rich variety of  planetary systems and protoplanetary disks. However, our incomplete knowledge of the processes of planet formation is limiting our ability to establish  a link between the two. In particular, the morphological evolution of protoplanetary disks is not fully understood. While the transition from a full gas-rich disk to a planetary system associated with a debris disk \citep[e.g.,][]{Williams2011} is a corroborated theory, there is no general consensus on the intermediate stages. The taxonomy of protoplanetary disks is the key to obtaining new insights into their evolutionary paths. 

A classification reflecting the \textit{radial} evolution of disks was proposed by \citet{Strom1989}. They noticed that some disks show a diminished near- to mid-IR flux and ascribed it to the rapid dissipation of the inner dusty material. A more recent classification by \citet{Meeus2001} involves the \textit{vertical} evolution of disks. They showed that the mid-/far-IR excess in the spectral energy distribution (SED) of intermediate-mass stars can either be fit by a power-law continuum or requires an additional cold blackbody. These two categories, named Group II and I, respectively (hereafter GII and GI), were originally thought to represent two distinct disk geometries, flat and flared disks, due to different stages of the dust grain growth and consequent settling toward the disk mid-plane  \citep[see, e.g.,][]{Dullemond2004}. This process would act to decrease the illuminated surface resulting in a smaller amount of reprocessed light and thus in the transition from GI to GII.

\begin{table*}
      \caption[]{Properties of Group I and II disks from different observational techniques.}
         \label{Literature}
     $$ 
         \begin{tabular}{cccc}
            \hline
            \hline
            \noalign{\smallskip}
            \hspace{15mm} Technique \hspace{15mm} & Group I & Group II & \hspace{7mm} Illustrative reference \hspace{7mm} \\
            \hline
            \hline
            \noalign{\smallskip}
            Near-IR scattered light & Routinely detected & Mostly undetected & \citet{Garufi2014b} \\
            \hline
                     \noalign{\smallskip}
        Near-IR ro-vibrational lines & Hot, from larger radii & Cold, from smaller radii & \citet{vanderPlas2015} \\
        \hline
            \noalign{\smallskip}
            Mid-IR continuum & Routinely resolved & Typically unresolved & \citet{Marinas2011} \\
            \hline
            \noalign{\smallskip}            
        Silicate emission features & Often detected & Always detected & \citet{Meeus2001} \\
            \hline
                 \noalign{\smallskip}
        Far-IR emission lines & Routinely detected & Mostly undetected & \citet{Meeus2012, Meeus2013} \\
            \hline
            \noalign{\smallskip}
        Millimeter continuum & Routinely resolved & Mostly unresolved & \citet{Mannings1997} \\    
            \hline
            \noalign{\smallskip}            
         Stellar abundances & Fe, Mg, Si subsolar & Fe, Mg, Si solar-like & \citet{Kama2015} \\
            \hline
            \hline
         \end{tabular}
     $$ 

   \end{table*}

These radial and vertical classifications may be more correlated than  was initially thought. In fact, the powerful imaging techniques of the last decade have highlighted the high recurrence rates of large cavities in GI disks \citep[e.g.,][]{Honda2015}. Conversely, no GII disk has shown the presence of a \textit{large} (tens of AU) inner gap \citep[whereas the presence of \textit{small} gap appears possible; see][]{Menu2015}. Also, \citet{Kama2015} found a systematic depletion of refractory elements only in the photosphere of stars hosting a GI disk, and they linked this to the large-dust trap exerted by giant planets in a disk cavity. Another dissimilarity is the significantly larger emitting radius of the CO ro-vibrational lines from the inner $\sim$10 AU in GI compared to GII \citep{Banzatti2015, vanderPlas2015}, which points toward the presence of a cavity in the molecular gas as well.  This dichotomy could also explain why all GII show silicate features, whereas a large fraction of GI do not. In fact, \citet{Maaskant2013} argued that the lack of silicate features is due to an intrinsic depletion of dust material where this emission originates. 

GII disks are also more elusive than GI. The mid-IR emission of GII is not resolved, contrary to GI \citep{Marinas2011, Honda2015}. The same consideration may apply at millimeter wavelengths since no resolved observations of a GII have been seen so far (with one possible exception,  HD163296,  which is discussed in this work). Furthermore, \citet{Garufi2014b} have shown that GI are routinely detected in scattered light, whereas GII are only marginally detected or not detected. Far-IR emission lines of CO and OH are strong in GI and remain mostly undetected in GII \citep{Meeus2012, Meeus2013}. All these observations may suggest that GII disks are self-shadowed, i.e., the inner disk intercepts most of the stellar light and casts a shadow far out \citep[e.g.,][]{Dullemond2001}. This scenario is supported by the anti-correlation between the brightness of the inner and outer disk \citep[e.g.,][]{Acke2009}. 

All this said, it is clear that the idea of an evolution from GI to GII has to be revised. \citet{Currie2010} and \citet{Maaskant2013} proposed that the two groups represent different tracks in the evolution of disks, where the most recognizable geometrical evolution is either gap formation or disk settling. Table \ref{Literature} summarizes the properties of GI and GII disks from various techniques.

In this paper, we present new near-IR observations of some GII in scattered light. The sample is described in Sect.\,\ref{Sect_Sample}. The observations are performed in polarimetric differential imaging (PDI),  a technique that allows a substantial fraction of the scattered light from the $\sim$$\mu$m-sized dust grains in the disk surface to be imaged \citep[e.g.,][]{Apai2004, Quanz2011}. The PDI technique is based on the simple and powerful principle that  stellar light is mostly unpolarized, whereas the scattered light from the disk is to a large extent polarized. The observational setup and data reduction are described in Sect.\,\ref{Sect_Observations}, and the results of the new dataset in Sect.\,\ref{Results}. The new observations of GII are then used to perform a taxonomic analysis of a larger sample of protoplanetary disks in Sect.\,\ref{Sect_Taxonomy}. A consequent discussion on the nature of GI and GII is finally given in Sect.\,\ref{Sect_Discussion} and Sect.\,\ref{Sect_Conclusions}.


\section{ Sample} \label{Sect_Sample}

The classification of GI and GII is not univocal. Contrary to the original Meeus classification, authors have defined multiple methods based solely on photometry which leave the partition of sources substantially unaltered. For example, \citet{vanBoekel2003} used the $m_{12}-m_{60}$ color in relation to five near-IR magnitudes. A more direct approach is to use the $[30\, \mu {\rm m}/13.5\, \mu {\rm m}]$ ratio since at these wavelengths the impact of spectral features of silicates and hydrocarbons is minimized \citep{Acke2009}. A strong correlation with the Meeus classification exists \citep{Maaskant2013} and, in this context, the transition between GI and GII may lie at $[30/13.5]=2.1$ \citep{Khalafinejad2016}. Also, \citet{Garufi2014b} found a correlation for this ratio with the amount of scattered light. We note that the set of thermochemical models by \citet{Woitke2016} showed that the [30/13.5] continuum ratio is significantly affected by two disk parameters, i.e.,\,the flaring angle and the inner radius. This degeneracy is discussed in Sect.\,\ref{Sect_Taxonomy}.

\begin{figure*}
  \centering
 \includegraphics[width=8.7cm]{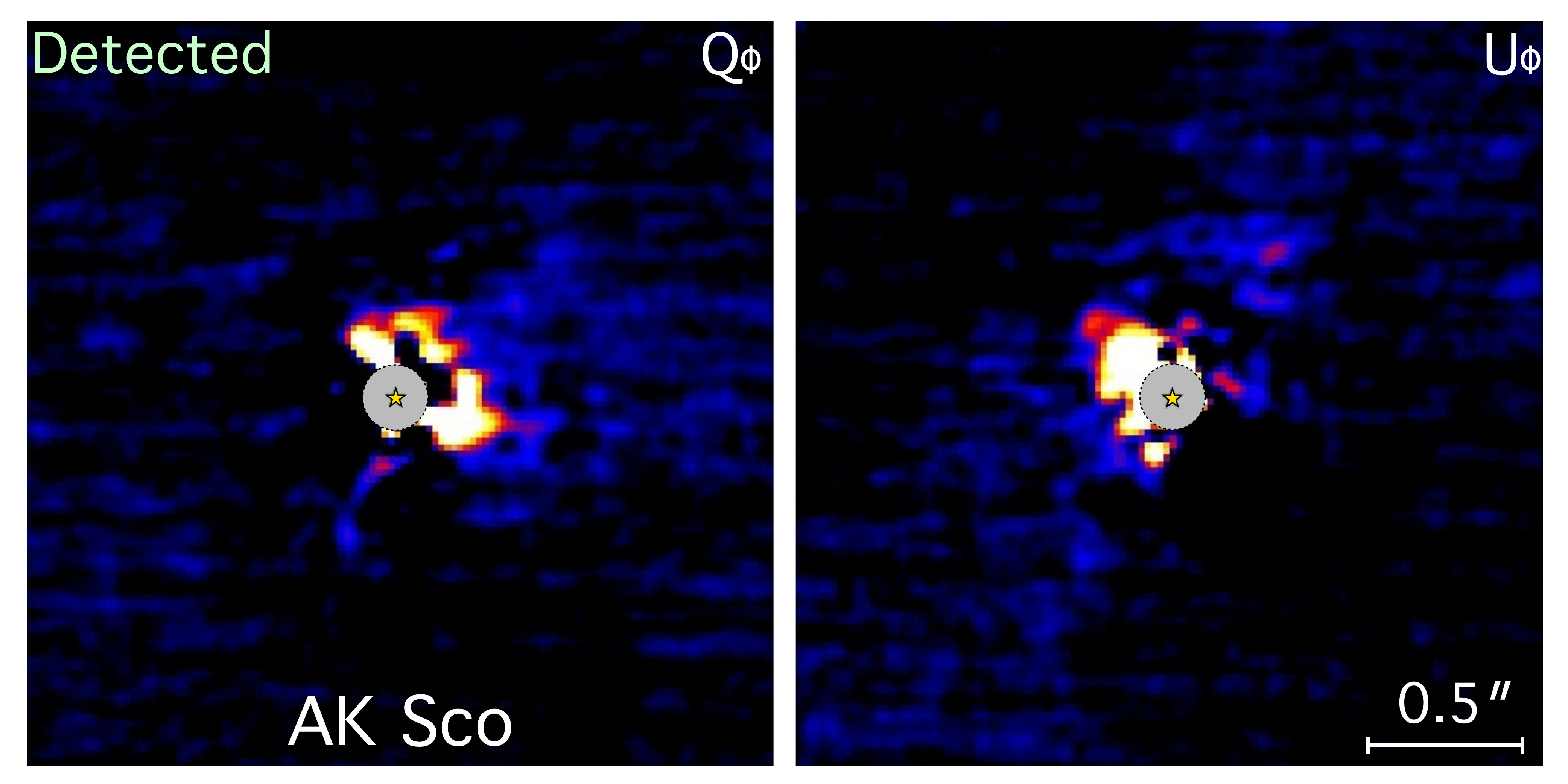} \hspace{7mm} 
 \includegraphics[width=8.7cm]{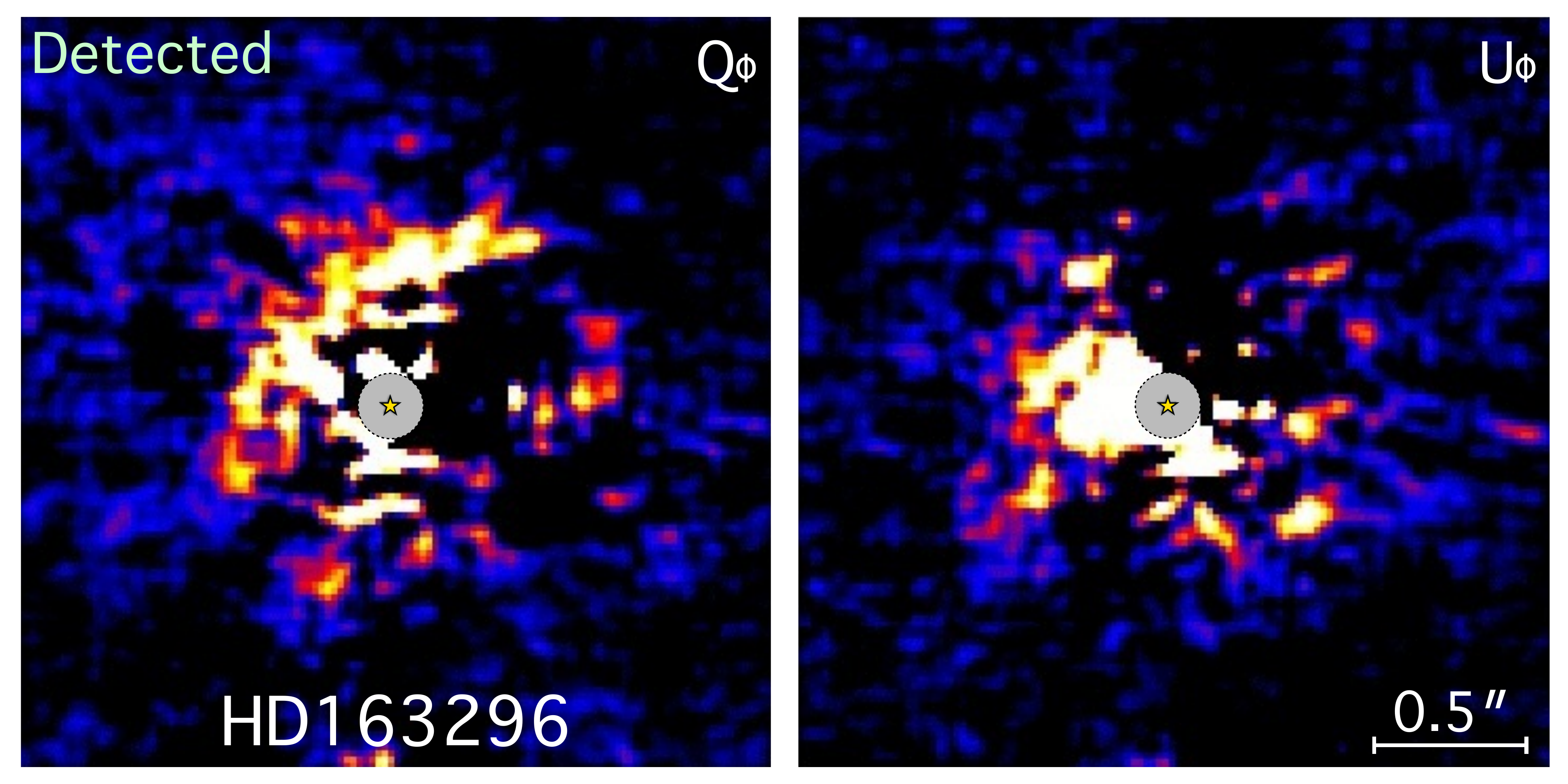} 
 \includegraphics[width=8.7cm]{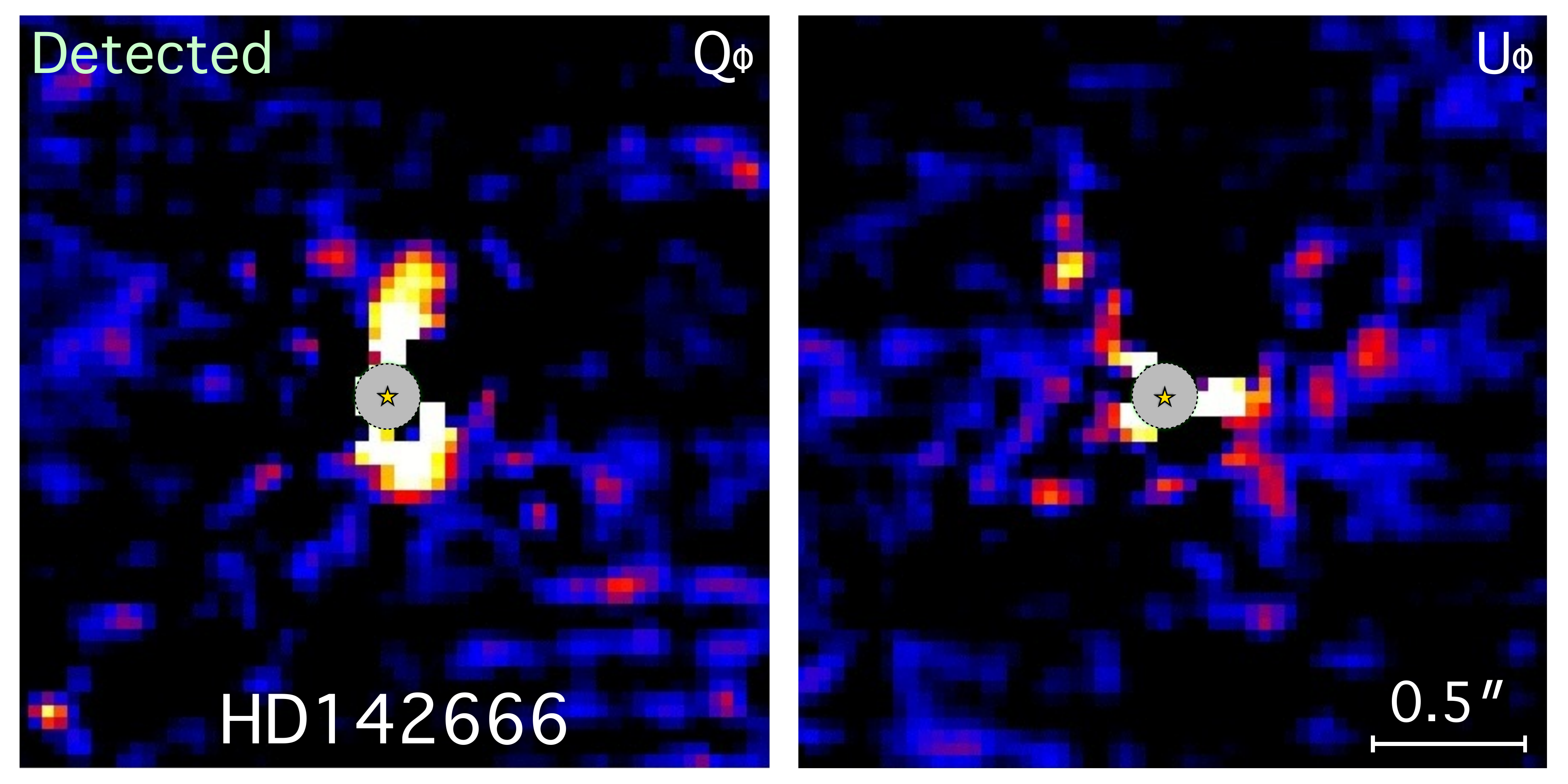} \hspace{7mm}
 \includegraphics[width=8.7cm]{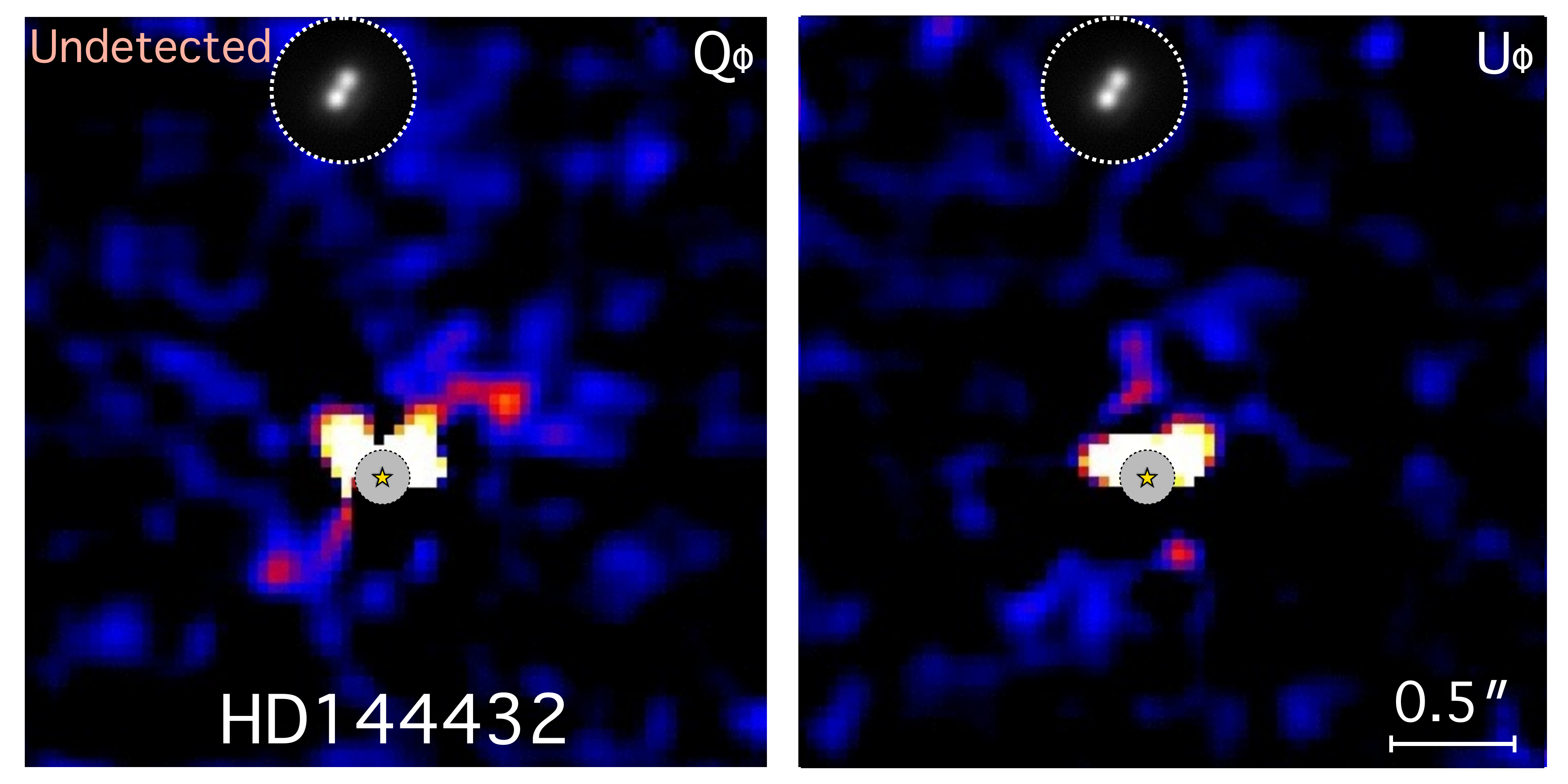}
 \includegraphics[width=8.7cm]{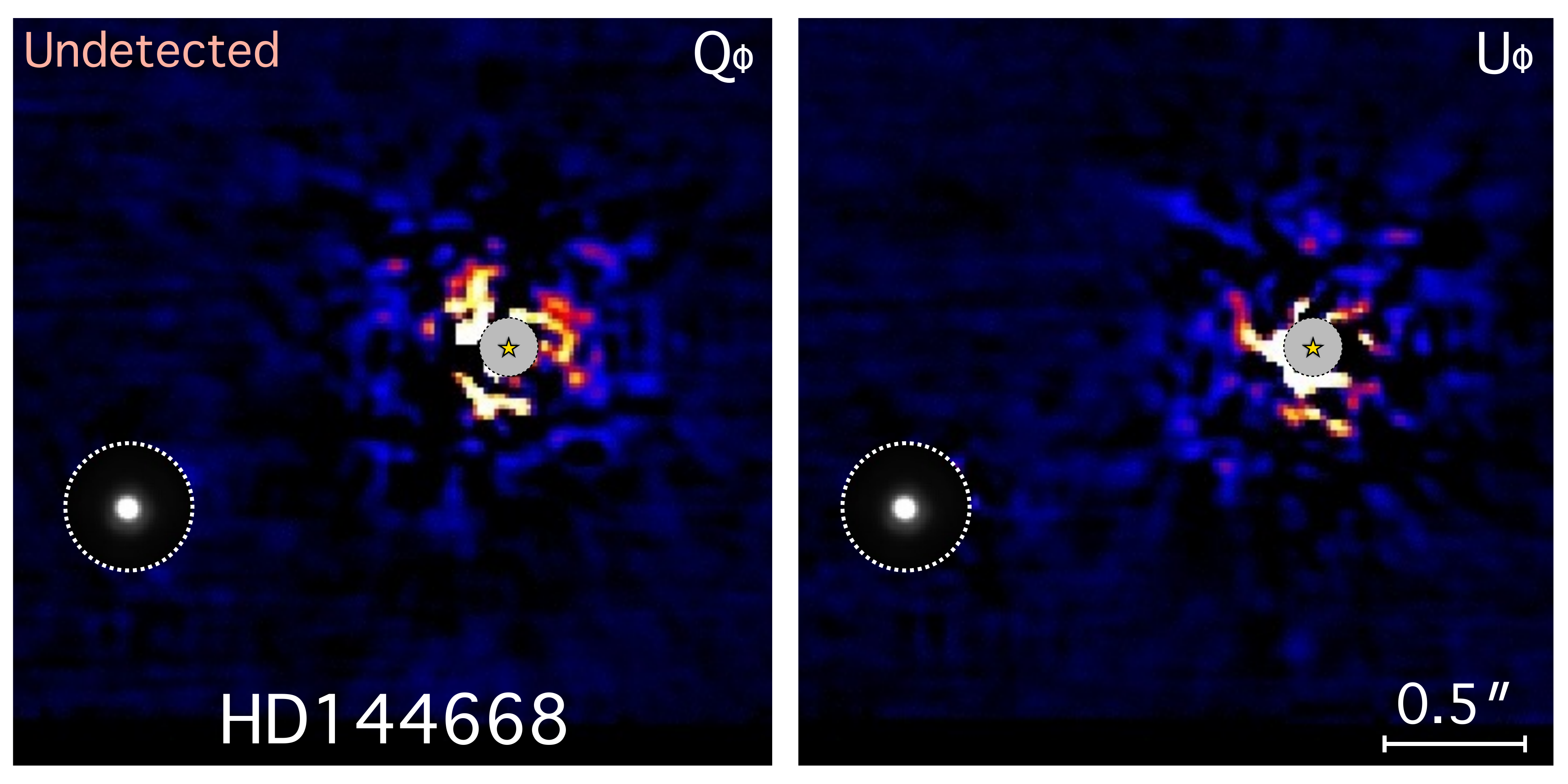} \hspace{7mm}
 \includegraphics[width=8.7cm]{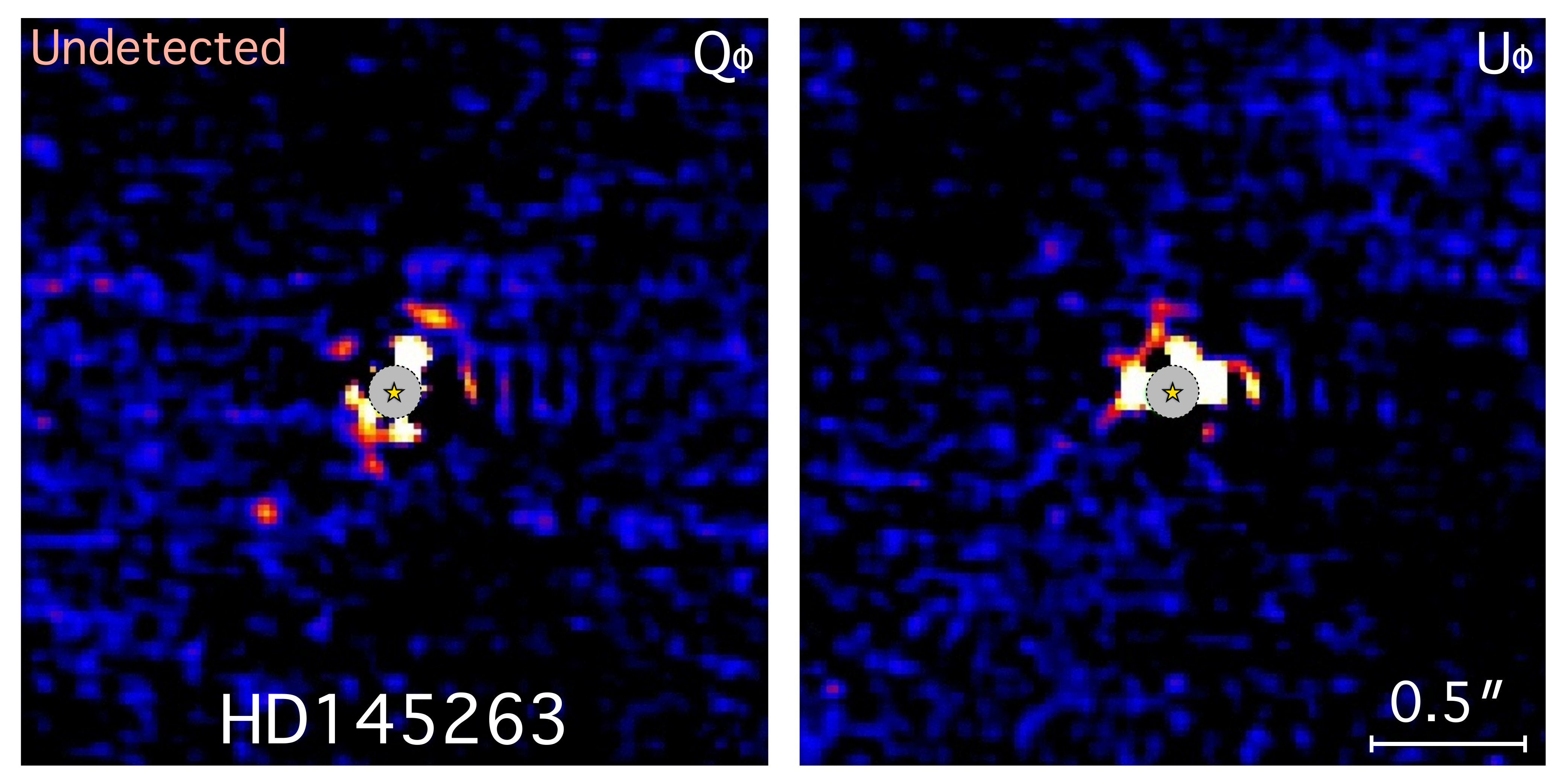}
  \caption{Polarized light imagery of the sample. For each object, the $Q_\phi$ image is shown to the left and the $U_\phi$ to the right. All pairs of images have the same linear color stretch and are \textit{not} scaled by the squared distance. The main stars are at the center of the $\sim$0.1$\arcsec$  green circles. When stellar companions are present, they are displayed with their $I$ image in an inset circle. North is up, east is left. }
          \label{Imagery}
  \end{figure*}

We used the [30/13.5] criterion to select the targets studied in this work. We chose sources with [30/13.5] around the above-mentioned threshold of 2.1.  The sample consists of six objects; they are briefly described below in order of mid-IR ratio:

\begin{itemize}

\item HD144668 (or HR5999, $[30/13.5]=1.0$) is an A5 star \citep{vandenAncker1997} forming a visual binary with a T Tau star at 1.4\arcsec \ \citep{Stecklum1995}. Optical emission lines suggest that the disk is close to edge-on \citep{Perez1993}. The mid-IR emission from the source is mostly unresolved, even though a faint extended signal is detected up to 95 AU to the N and S by \citet{Marinas2011}. 

\item HD142666 ($[30/13.5]=1.5$) is an A3/A8 star \citep{Blondel2006} that is thought to be isolated. The circumstellar disk has an inner radius slightly larger than the dust sublimation radius at sub-AU scale \citep{Schegerer2013, Menu2015}. 

\item HD144432 ($[30/13.5]=1.8$) is the primary A8 star \citep{Sylvester1996} of a triple system \citep{Mueller2011}. Sub-mm observations of the CO emission lines reveal a ${\sim 60}$ AU  disk which is $\sim 45\degree$ inclined \citep{Dent2005}. Similarly to HD142666, near-IR interferometry \citep{Chen2012} hints at the presence of a small-scale cavity.

\item HD163296 ($[30/13.5]=2.0$) is the best-known object of the sample. The disk around the isolated A1 star \citep{Mora2001} has been systemically imaged at (sub-)mm \citep[e.g.,][]{Isella2007} and near-IR wavelengths \citep[e.g.,][]{Grady2000}. ALMA observations \citep{Guidi2016} revealed an excess in the mm emission in proximity to the radial location of the CO snowline \citep{Qi2015} and to a ring visible in scattered light \citep{Garufi2014b}. This excess may be due to an increase in dust density caused by a local dust trapping. 

\item HD145263 ($[30/13.5]=2.0$) is an F0 star \citep{Smith2008} with a disk exhibiting an IR excess of intermediate amount between a young and a debris disk \citep{Honda2004}. 

\item HD152404 (or AK Sco, $[30/13.5]=3.3$) is a F5+F5 close binary \citep[with a separation of 0.16 AU,][]{Anthonioz2015}. It is classified as a GII source based on its far-IR excess. However, according to the mid-IR criterion it is a GI source whose  [30/13.5] ratio is among the lowest known for GI. ALMA images show a relatively extended disk \citep{Czekala2015} which is also resolved with SPHERE near-IR observations, revealing a double-wing structure \citep{Janson2016} that may imply the existence of a gap or of spiral arms.

\end{itemize}

The sample studied in this work is complemented by 11 additional B-to-F type stars with near-IR PDI images available from the literature. A description of the full sample used in Sect.\,\ref{Sect_Taxonomy} can be found in Appendix \ref{Appendix_sample}.


\section{Observations and data reduction} \label{Sect_Observations}

\begin{table}
      \caption[]{Summary of observations. Columns are: night number (see text), source name, detector integration time (sec) multiplied by number of integrations and by number of exposures, total integration time (sec), and DIMM seeing during the observation. We note that $t_{\rm exp}={\rm DIT}\times{\rm NDIT}\times{\rm NE}\times 4$ with 4 being the halfwave plate positions. }
         \label{Observations}
     $$ 
         \begin{tabular}{ccccc}
            \hline
            \hline
            \noalign{\smallskip}
            Night & Source & DIT(s)$\times$NDIT$\times$NE & $t_{\rm exp}$ (s) & Seeing \\
            \hline
            \noalign{\smallskip}
     1 & HD144432 & 0.3454$\times$148$\times$20 & 4,090 & 1.0\arcsec -1.6\arcsec \\
     \hline
     \noalign{\smallskip}
     1 & HD163296 & 0.3454$\times$148$\times$20 & 4,090 & 1.1\arcsec -2.5\arcsec \\
     \hline
     \noalign{\smallskip}
     1 & HD152404 & 2$\times$40$\times$24 & 7,680 & 1.2\arcsec -2.4\arcsec \\ 
     \hline
     \noalign{\smallskip}
     2 & HD144668 & 0.3454$\times$130$\times$24 & 4,310 & 0.8\arcsec -1.3\arcsec \\
     \hline
     \noalign{\smallskip}
     2 & HD142666 & 0.5$\times$112$\times$20 & 4,480 & 0.8\arcsec -1.3\arcsec \\
     \hline
     \noalign{\smallskip}
     2 & HD145263 & 2$\times$25$\times$20 & 4,000 & 0.8\arcsec -1.4\arcsec \\         
            \hline
            \hline
         \end{tabular}
     $$ 
   \end{table}

The six new sources of this work were observed over two nights (\mbox{22-23} July 2015) with the Adaptive Optics(AO)-assisted NAOS/CONICA instrument \citep[NACO,][]{Lenzen2003, Rousset2003} at the Very Large Telescope (VLT) in polarimetric differential mode (PDI). The observing strategy followed that used in the works with NACO by \citet{Quanz2013b}, \citet{Garufi2013}, and \citet{Avenhaus2014a}. In PDI with NACO, the stellar light is split into two beams containing orthogonal polarization states by a Wollaston prism. A rotatable half-wave plate provides a full cycle of polarization state at 0$\degree$, 45$\degree$, 90$\degree$, and 135$\degree$. At the time of the observations, NACO was fixed in the S13 objective. The small scale of this camera (13.27 mas/pixel) and the high background level caused by insufficient shielding in the instrument impeded us from performing optimal observations. Furthermore, one of the camera quadrant{s} that we partly used showed a non-static noise across the detector rows that could not be completely removed during the data reduction. 

All targets were observed in the $K_{\rm S}$ band for a total time of 1.12 to 2.13 hours per source (see Table \ref{Observations}). The sources had an average airmass of $\sim$1.2. We slightly saturated the stellar PSF in the inner few pixels to obtain a better signal-to-noise ratio. We also dithered the stellar position on the detector to evaluate any artifacts. Both nights were affected by a highly variable seeing ($0.66\arcsec-2.68\arcsec$) with the intermittent passage of thin cirrus. 

The data reduction was performed following the method outlined by \citet{Avenhaus2014a}. Apart from the standard cosmetic steps (dark current subtraction, flat fielding, bad pixel correction), this method consists in upscaling the image before determining the star position to achieve a sub-pixel accuracy, extracting the two beams with perpendicular polarization states, and equalizing the flux contained in annuli from the two beams to attenuate the instrumental polarization. The final images are produced by combining the beams from the full polarimetric cycle into the parameters $Q_\phi$, $U_\phi$, $P$, and $I$ \citep[see][for details]{Schmid2006}\footnote{{The pair of parameters ($Q_\phi$, $U_\phi$) is referred to as ($Q_r$, $U_r$) in the reference paper. Other authors have also used the nomenclature ($P_\perp$, $P_\parallel$) and ($Q_T$, $U_T$).}}. The images shown in the paper are binned by 1.5 with respect to the original pixel size to reduce the shot-noise and are smoothed with a Gaussian kernel of 0.04\arcsec (approximately half of the instrument PSF).


\begin{figure}
 \centering
 \includegraphics[width=7.5cm]{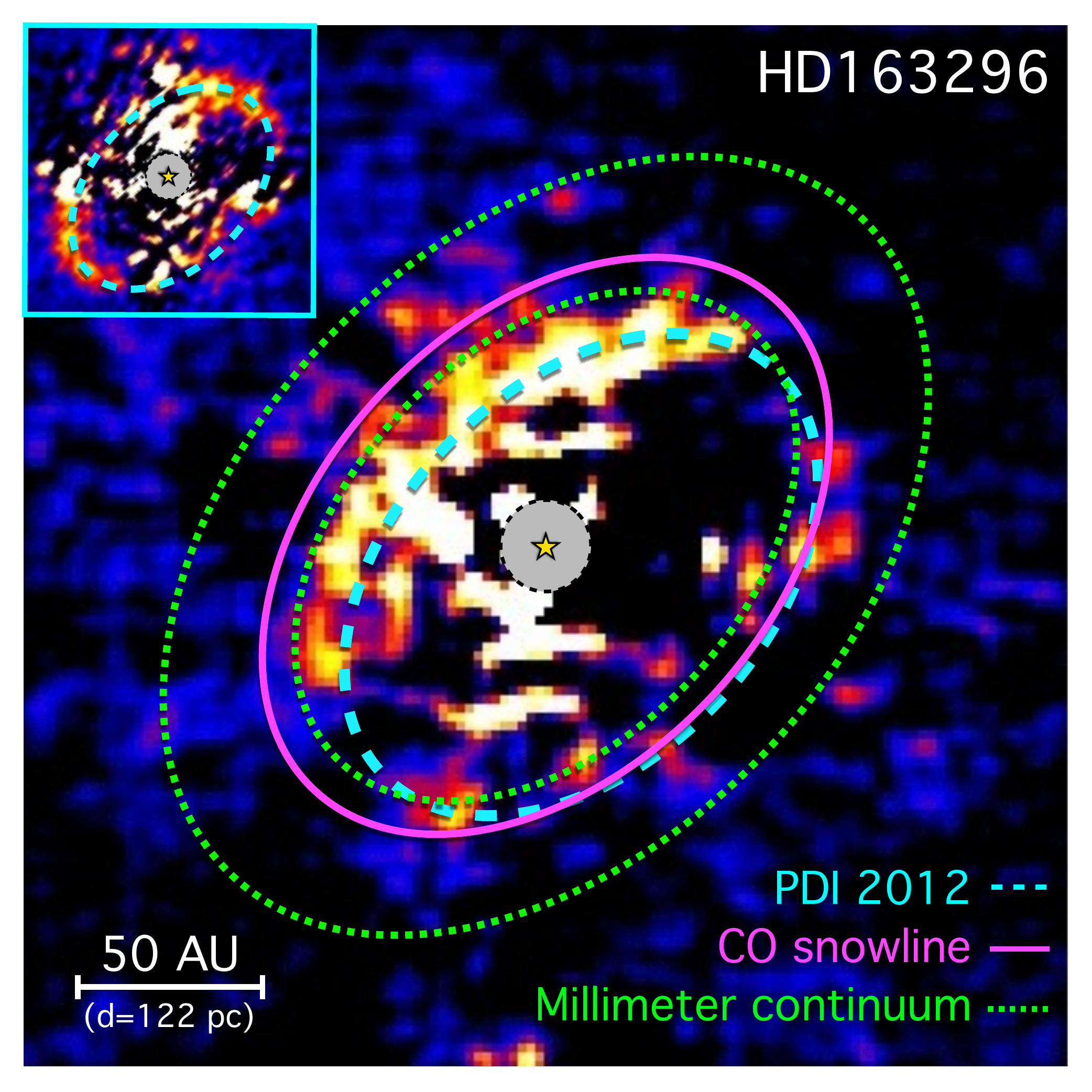} 
  \caption{PDI image of HD163296 from this paper compared to other works. The cyan dashed line indicates the peak intensity of the ring detected in scattered light by \citet{Garufi2014b} and shown in the inset image. The violet solid line lies at the CO snowline \citep{Qi2015} taking into account the disk inclination and keeping the star in the center. The green dotted lines are obtained similarly from the peak intensity of the two innermost rings revealed at 1.3 mm by \citet{Isella2016}. The star is at the center of the gray circle. North is up, east is left.} 
          \label{HD163296}
  \end{figure}

\section{ New polarized images} \label{Results}
The final PDI images of the six sources are shown in Fig.\,\ref{Imagery}. Even though these observations are non-coronagraphic, we consider the inner $\sim0.1\arcsec$ around the central star unreliable and thus we masked out the region. This is motivated by both the smearing effect due to the finite PSF resolution \citep[see][]{Avenhaus2014b} and by the suboptimal AO correction of these observations.  

\subsection{Results}

\hspace{4mm} \textbf{AK Scorpii.} The disk around AK Sco is clearly detected in the $Q_\phi$ image. Two bright lobes are seen along the NE-SW direction and their signal can be detected inward down to the innermost reliable distance from the star ($\sim 0.07\arcsec$). In the NW quadrant, the signal is detected in the form of an arch connecting the two lobes. No significant signal is detected in the SE quadrant. The $U_\phi$ also shows a strong signal along the NE axis in correspondence with one $Q_\phi$ lobe. 

\textbf{HD163296.} The top right panel of Fig.\,\ref{Imagery} shows the detection of the ring-structure around HD163296. The emission is maximized to the north, while it is very marginal to the south. Two symmetric minima are seen along the SE-NW direction. Since these minima correspond to the strong AO spots of these observations, they should not be trusted. Inside the ring, we cannot reveal any coherent disk structures. Similarly to AK Sco, the signal in the $U_{\phi}$ is very strong, with a positive branch on one side of the major axis and a negative branch on the other. 

\textbf{HD142666}. The $Q_\phi$ of this star reveals a relatively strong signal to the north and to the south that is persistent across individual frames. This notion and the different brightness and distribution of the $U_\phi$ image suggest that the signal detected in the $Q_\phi$ is actually a signal from the circumstellar disk. The signal from $Q_\phi$ extends inward at least down to the innermost reliable radius. 

\textbf{HD144432}. The binary companions of HD144432 are easily detected in the intensity image to the north of the main star. In particular, the center of the binary system lies at $r=1.48\arcsec \pm 0.01\arcsec$ with P.A. $=5.48\degree \pm 0.19\degree$. Thus, the system has moved counterclockwise from 2005 to 2015 by $1.17\degree$ \citep[based on the astrometric analysis by][]{Mueller2011}. Also, the pair of companions has moved relative to each other by $\Delta r= -0.016\arcsec$ and $\Delta$P.A.$=-57.57\degree$. The $Q_\phi$ and $U_\phi$ are comparable in both morphology and intensity. An extended feature is visible from $Q_\phi$ across the SE-NW axis. This is persistent in many individual frames and absent in the $U_\phi$ image. However, the vertices of the feature match the location of the AO spots in the intensity image. Thus, we consider the signal as spurious.

\textbf{HD144668}. The intensity image of HD144668 also reveals the presence of a companion, at $r = 1.46\arcsec \pm 0.01\arcsec$ and P.A. $=111.88\degree \pm 0.25\degree$. Thus, the orbital motion from 1992 \citep[$r=1.459\arcsec$ and P.A. $=110.96\degree$,][]{Stecklum1995} is still very marginal. A very uneven signal is seen in both the $Q_\phi$ and the $U_\phi$ caused by a very variable PSF across the observations, and the presence of scattered light cannot be inferred.  

\textbf{HD145263}. Both the $Q_\phi$ and the $U_\phi$ show only a very marginal signal close to the star with comparable brightness. Thus, no polarized light is detected from this object. 

\subsection{Interpretation}
Three of the six disks in the sample are detected. The strongest signal is revealed around AK Sco, which is the source with the highest [30/13.5] ratio in the sample (see Sect.\,\ref{Sect_Sample}). The morphology of this signal is to first order consistent with that of the SPHERE image by \citet{Janson2016}, obtained with angular differential imaging (ADI). The only significant difference is the presence in the PDI images of the NW arch, which may represent the far side of a full ring that remained undetected in the ADI image. In fact, \citet{Garufi2016} showed that the process of ADI acts to damp the disk emission from the minor axis and thus that an azimuthally symmetric feature can be seen as a double-wing structure aligned with the major axis. We defer further considerations on the disk geometry to a forthcoming SPHERE paper. 

The presence of signal in the $U_{\phi}$ image of both AK Sco and HD163296 is qualitatively consistent with the deviation from tangential scattering from inclined disks (these disks are $\sim70\degree$ and $\sim45\degree$ inclined), which acts to redirect part of the polarized signal from the $Q_{\phi}$ to the $U_{\phi}$ image \citep{Canovas2015}. We emphasize that the signal close to the star in $U_{\phi}$ may appear stronger than from other works partly because these images, for consistency with the rest of the dataset, are \textit{not} scaled with the squared distance from the star. In any case, a partly instrumental contribution cannot be ruled out because of the NACO cross-talk effect between the Stokes parameters \citep{Witzel2010}.

In Fig.\,\ref{HD163296} we show the polarized image of HD163296 and compare it to that obtained with the same mode in 2012 (see inset image) by \citet{Garufi2014b}. The evident difference between the two is largely due to the spurious signal present in both datasets. For example, the northern region (the brightest in the 2015 dataset) was mostly unaccessible in 2012 because of the presence of a strong artifact. In any case, the radial location and the apparent flattening of the ring remains to first order unchanged between the two epochs (see cyan dashed lines). This demonstrates that the ring in scattered light is due to a persistent disk morphology rather than  a transient shadow from the inner disk \citep[as proposed by][]{Garufi2014b}. This finding reinforces the spatial connection with the excess in the continuum emission at 850 $\mu$m shown by \citet{Guidi2016} and with the location of the CO snowline (see violet solid line) as inferred by \citet{Qi2015}. Recent ALMA images \citep{Isella2016} have revealed the presence of three rings in the millimeter continuum that are indicated in Fig.\,\ref{HD163296} by the green dotted lines. \citet{Guidi2016} discussed a scenario where dust trapping is favored at the CO iceline and results in an increased dust surface density. This should also have an impact on the disk surface to allow the detection of scattered light from a disk elsewhere undetected. In particular, the PDI ring is nearly co-spatial with the innermost ALMA millimeter ring at $\sim$80 AU. 

Finally, the signal detected around HD142666 points toward an inclined disk with the major axis oriented in the north-south direction. This geometry is consistent with what was inferred for the inner disk by \citet{Vural2014}, i.e.,\ $i \sim 50\degree \ {\rm{and}\ P.A. \sim 170\degree}$. The presence of a signal as close to the center as $\sim 0.07\arcsec$ rules out the existence of any cavity larger than $\sim$10 AU. More importantly, the abrupt outward decrease in signal at $0.4\arcsec$ is most likely not due to the sensitivity. Thus, these observations are consistent with a rather compact disk of $\sim$60 AU in size. 

The non-detection of the other disks can be due to many different factors (e.g.,\, self-shadowing, deficit of scattering particles, dust properties). These possibilities are explored in the broader context of the dichotomy between GI and GII throughout the paper.

\section{Taxonomy of Group I and Group II} \label{Sect_Taxonomy}
In this section, we investigate the brightness of a number of GI and GII disks in scattered light and explore connections with the disk properties known from the literature. The sample consists of ten GI and seven GII disks that have been observed in near-IR PDI over the last five years with either VLT/NACO or VLT/SPHERE. The full sample is described in Appendix \ref{Appendix_sample}.

Measuring the amount of scattered light from a sample of disks in a consistent way is not an easy task. In fact, the stellar brightness, the distance of the source, and the disk inclination significantly alter the intrinsic amount of light that we detect. However, it is possible to compute the polarized-to-stellar light contrast along the disk major axis to elude their influence. By doing this, the polarized-to-stellar light contrast (referred to as contrast hereafter) is a direct measurement of the capability of the disk surface to scatter photons, which in turn depends on the disk geometry and the dust properties. A detailed description on the calculation of the contrast can be found in Appendix \ref{Appendix_contrast}.

\begin{figure*}
 \centering
 \includegraphics[width=15cm]{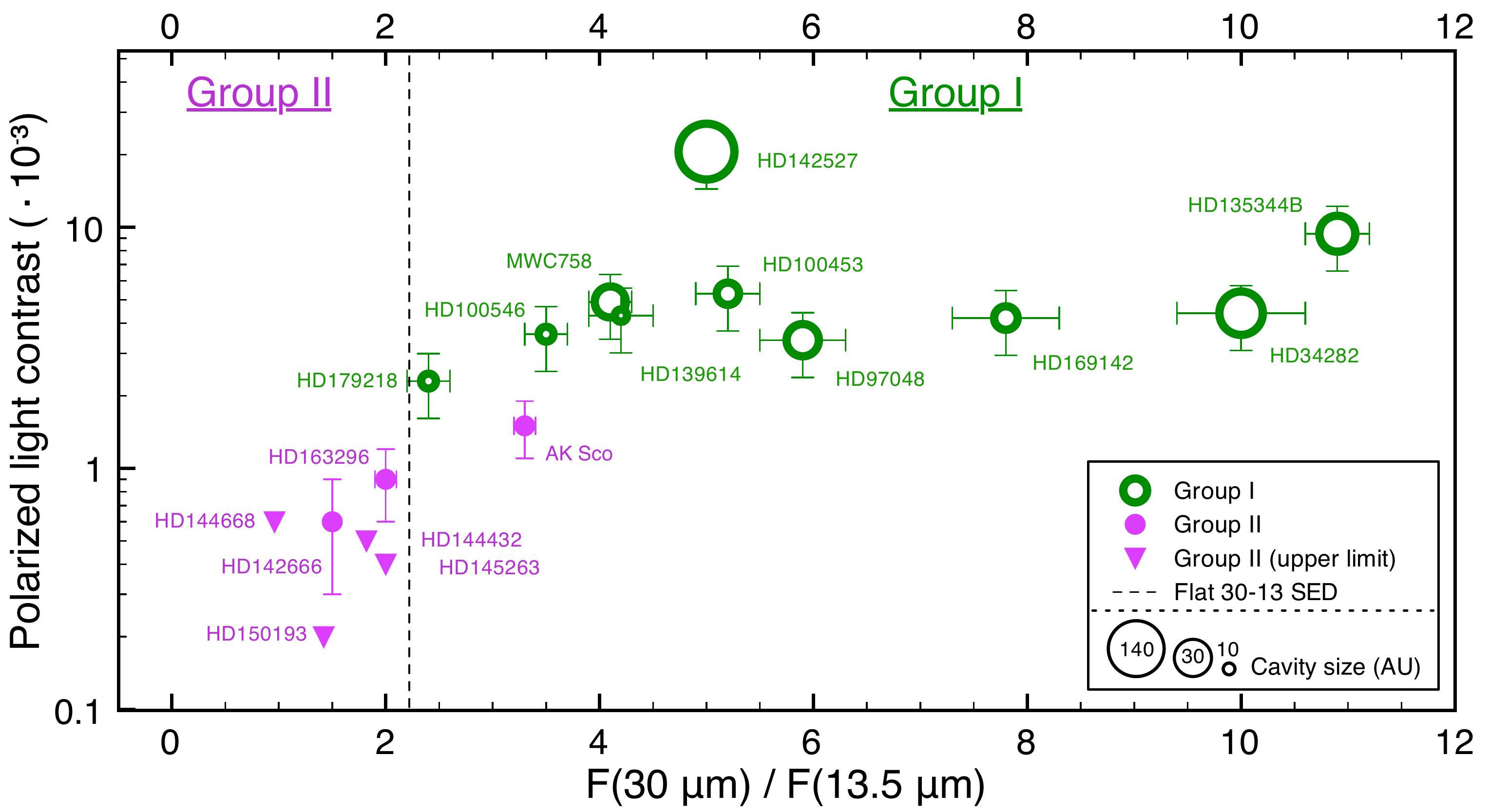} 
  \caption{Polarized-to-stellar light contrast for all the sources in the sample (see Appendix \ref{Appendix_sample}) compared with the flux ratio at 30 $\mu$m and 13.5 $\mu$m. GI disks are plotted in green,  GII in purple. The disk cavity, where known and as taken from different datasets (see text), is indicated by a gap in the symbol, proportional to the cavity size with dynamic range from 5 AU to 140 AU. The dashed line indicates the ratio corresponding to a flat SED, obtained from $30\div13.5=2.2$. The ratios are from \citet{Acke2010}, while the contrasts are from this work, as explained in Appendix \ref{Appendix_contrast}.} 
          \label{Contrast_color}
  \end{figure*}

The measured contrast for all the targets in the sample is shown in Fig.\,\ref{Contrast_color}. From the plot, it is clear that GI disks are systematically brighter than GII disks in scattered light. With few exceptions, all GI disks have comparable contrast. It is also evident that the dichotomy between GI and GII can be expressed in terms of the presence or absence of a disk inner cavity. In the plot, we also show the cavity sizes constrained by millimeter images (where available), PDI images (when the cavity is detected), or SED fitting (see Appendix \ref{Appendix_sample} for details). We defer the discussion on discrepant estimates from these techniques to Sect.\,\ref{Discussion_gapped}. Interestingly, the disks with larger cavities ($R \gtrsim 15$ AU) have higher [30/13.5] ratios. Qualitatively, this is a consequence of the fact that this ratio depends on  the disk flaring angle and on the radial location of the disk inner edge \citep[e.g.,][]{Woitke2016}. In fact, for disks with no cavities or small cavities ($R \lesssim 15$ AU, \mbox{[30/13.5] $<$ 4}) a possible trend is seen in the diagram since both the ratio and the contrast are primarily affected by the flaring angle. Three of the four non-detections \citep[from this work and][]{Garufi2014b} are inconsistent with the trend. For sources with cavities larger than $\sim$15 AU, the relation is no longer present possibly because the ratio is mostly affected by the deficit of inner material. This idea is supported by the notion that the four sources in the plot with the highest [30/13.5] ratios are the only ones in the sample that do not show silicate features. In fact, the most plausible explanation for this lack is the deficit of material where this emission originates \citep[e.g.,][]{Maaskant2013, Menu2015, Khalafinejad2016}.

Figure \ref{Contrast_color} can be also used to obtain a first-order estimate of the polarized flux to be expected from new observations of disks. Disks with [30/13.5] $>$ 4 will be easily detected by new-generation instruments like SPHERE or GPI, whereas observations of disks with lower ratios require deeper integrations (longer than 1 hour).

We now investigate how the stellar and the disk properties of these sources relate with the scattered light contrast.

\subsection{Stellar properties}
To investigate whether the polarized contrast is related to the stellar properties, we compare our contrasts with the effective temperature of the stars (see Fig.\,\ref{Contrast_stellar}a). It turned out that GI and GII disks in  the sample are uniformly distributed across stellar temperature and mass. There is an accidental selection valley in the sample:  6 stars are warmer than 9,000 K and 11 colder than 8,000 K (note the discontinuity in the x-axis). In the figure, we also label those disks that show peculiar structures in scattered light, namely rings or spirals\footnote{For most disks, this classification is obvious. Two cases are subject to interpretation: HD100546, showing wrapped arms that may resemble rings \citep{Garufi2016}, and HD142527, showing a disk wall with multiple spiral arms outward of it \citep{Canovas2013}. We do not classify these objects here because  their  natures are different from the nominal ring-like disks \citep[e.g., HD97048,][]{Ginski2016} or symmetric spiral-like disks \citep[e.g., HD135344B,][]{Garufi2013}}. The ring-like disks in the sample are found predominantly around B and early-A  stars ($T_{\rm eff} > 7500$ K) while spiral-like disks are all found around late-A and F stars (\mbox{$T_{\rm eff} < 7700$ K}). 

In Fig.\,\ref{Contrast_stellar}b, we show the radial range of detection of disks, from both scattered light and millimeter continuum, as a function of the stellar temperature. Disks around early stars are routinely detected on larger scales (on average $\sim$340 AU vs $\sim$170 AU). There is no correlation between the cavity size and the stellar type. Two of the three GII disks detected are significantly smaller than the GI disks. The only large GII is HD163296, which may differ in many other aspects from the other GII disks (see Sect.\,\ref{Discussion_types}). Interestingly, three of the four non-detections have a companion at a projected distance of 100-200 AU. The only two GI disks with an outer companion are those with smaller detected extent, suggesting that the outer disk truncation may play an important role (see Sect.\,\ref{Discussion_types}).

\subsection{Outer disk properties}
Spatially unresolved information on the outer disk structure can be obtained from the SED at wavelengths from the mid-IR to the millimeter regime. We compare some of these constraints with our contrast in Fig.\,\ref{Outer_disk_contrast}.

\subsubsection{Far-IR excess}
{We calculated the far-IR excess from 20 $\mu$m to 450 $\mu$m of all sources, following the method described by \citet{Pascual2016}}. This excess is plotted against the contrast in Fig.\,\ref{Outer_disk_contrast}a. We found a clear trend between the two quantities. This relation is due to the co-located origin of the far-IR thermal light and the near-IR scattered light, i.e.,\ the disk surface at tens of AU. The following linear regression is found to fit the data:
\begin{equation}
\frac{F(\rm FIR)}{F_*}=34.97 \times \phi_{\rm pol}
\end{equation} 
In words, this trend says that on average the amount of flux scattered (and polarized) by the disk in the near-IR is $2.86\%$ ($\equiv 1 \div 34.97$) of the value of the thermal far-IR flux. In particular, we found values spanning from {1.1}$\%$ to 5.9$\%$. These values are only upper limit{s} of the real polarized scattered/thermal energy budget since the far-IR is a global measurement (and thus affected by the disk inclination), whereas our contrast is a local measurement (not affected by the disk inclination). We did not find any correlations for the contrast with any far-IR photometry (at 70 $\mu$m, 100 $\mu$m, and 160 $\mu$m), indicating that the disk flaring angle cannot be properly estimated from a single waveband.

\begin{figure*}
 \centering
 \includegraphics[width=9cm]{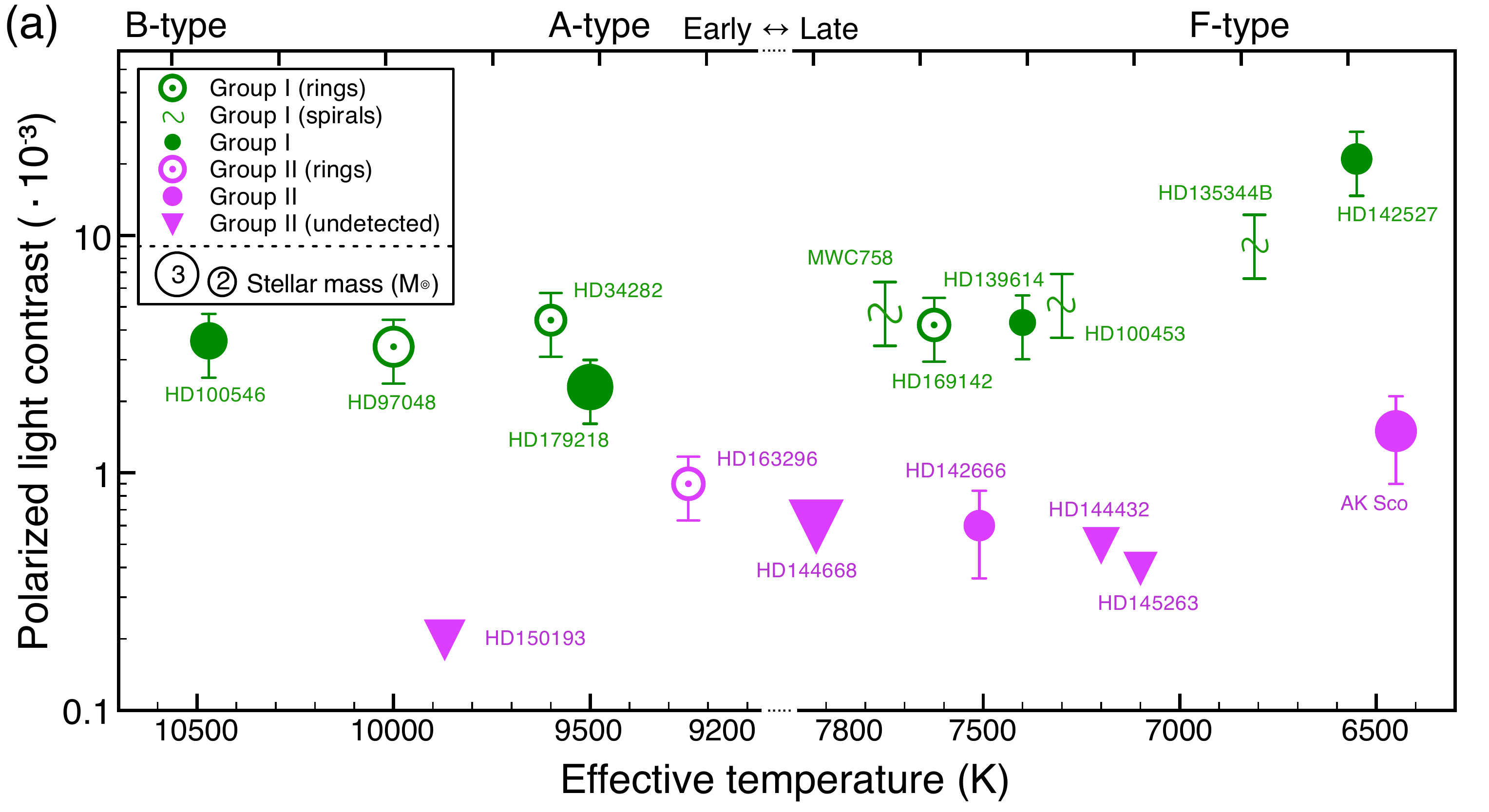} 
  \includegraphics[width=9cm]{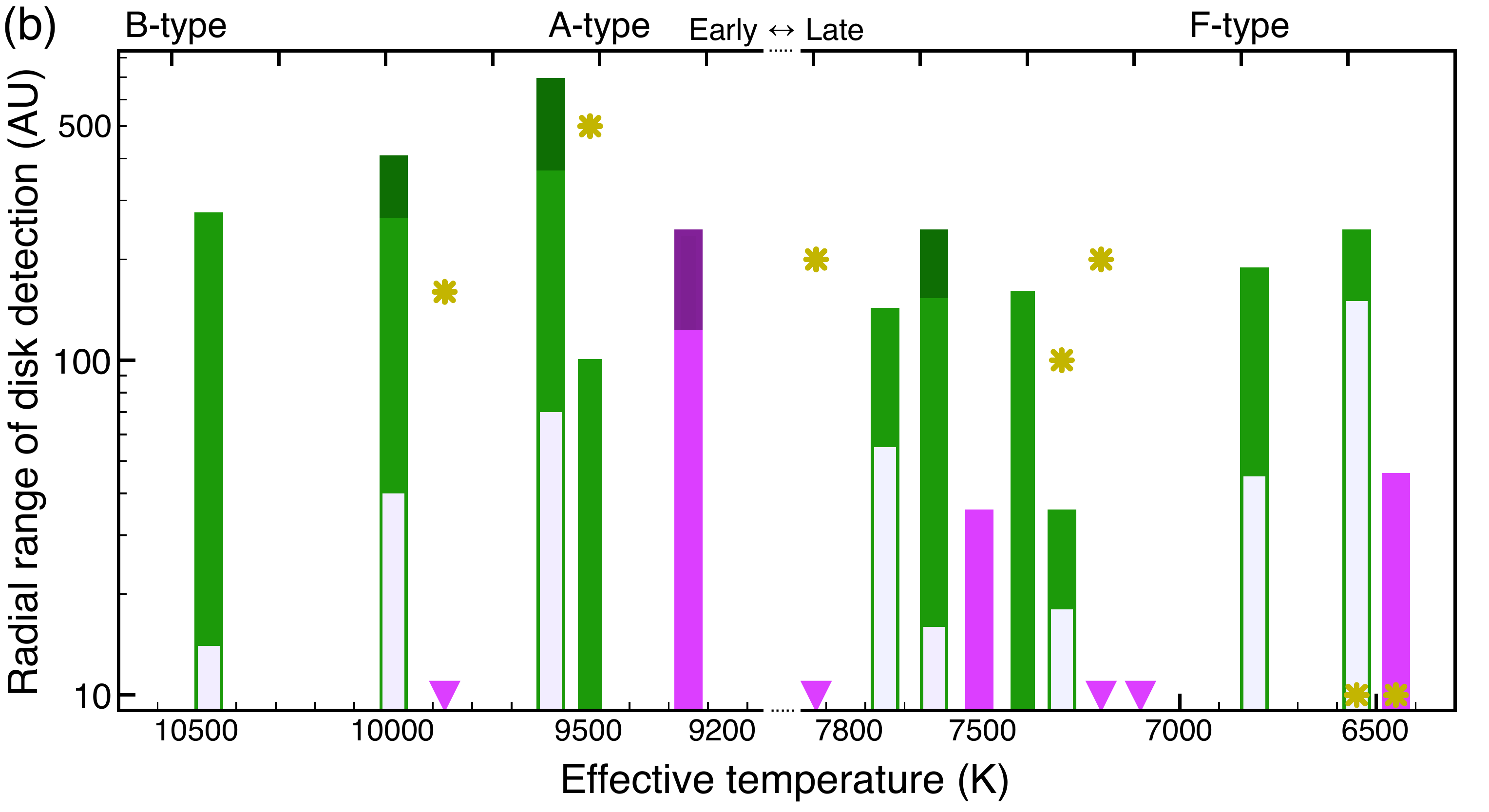} 
  \caption{Disk properties compared to stellar properties. \textbf{(a)}: Polarized-to-stellar light contrast compared to the stellar effective temperature. Note the discontinuity on the x-axis. The symbol size indicates the stellar mass with dynamic range between 1.6 M$_{\odot}$ and 3.2 M$_{\odot}$. \textbf{(b)}: Radial range of detection in PDI, compared to the stellar effective temperature. When continuum millimeter imaging reveals a larger radius, this is indicated by the darker  areas at the top of the bars. The white areas indicate disk cavities.  The yellow symbols give the projected distance of the companions. When these symbols are at the base of the bars, they indicate a binary system surrounded by the disk.} 
          \label{Contrast_stellar}
  \end{figure*}

{There are three significant outliers to the faint wing of the distribution. Even though HD150193 and HD144432 have a far-IR excess comparable to the other GII, their disks are not detected in scattered light \citep[this work and][]{Garufi2014b}, making the scattered/thermal flux ratio $< 0.6\%$ and $< 1.2\%,$ respectively.} A possible explanation is that the whole disk is less extended than the inner working angle of the PDI observations ($\lesssim 15$ AU). On the other hand, the non-detection of HD144668 is still consistent with the trend.

\subsubsection{Polycyclic aromatic hydrocarbons}
The gaseous disk can be traced by the mid-IR emission of polycyclic aromatic hydrocarbons (PAH). Direct imaging of PAH has shown that this emission can originate from the outer regions of disks \citep{vanBoekel2004, Lagage2006}. The emission from GI is typically stronger than from GII \citep[e.g.,][]{Acke2010}. In Fig.\,\ref{Outer_disk_contrast}b we show the PAH luminosity relative to the star as obtained by \citet{Acke2010} and compare it with the contrast. 

From the plot, the dichotomy for GI and GII is evident. Among the GI, the stellar temperature correlates with the PAH luminosity. All GI have prominent PAH emission. Among the GII, only three disks are detected and these are not sources with high stellar temperature. One GII, HD142666, even shows PAH brightness comparable to the GI. This is the most significant departure from the expected correlation between the PAH strength (to the first order, tracing the gas) and the contrast (tracing the dust). A possible explanation to these departures may derive from the new view that GI are gapped disks and GII are not, which is discussed in Sect.\,\ref{Discussion_evolution}.

\subsubsection{Millimeter flux}
The millimeter flux of the sources is compared to the contrast in Fig.\,\ref{Outer_disk_contrast}c. The fluxes at 1.3 mm obtained by multiple authors (see Appendix \ref{Appendix_sample}) have been scaled to a distance of 140 pc by means of the new GAIA measurements \citep{Gaia2016}, where available. Thus, the relative uncertainties are smaller than in previous works. As can be seen from the figure,  there is no clear trend between the millimeter flux and the polarized contrast. There is also no clear correlation with the stellar mass. Interestingly, all GII in the sample, except HD163296, are fainter in millimeter than the GI:  the former group has an average $F_{1.3 \rm mm}\simeq60$ mJy and the latter $F_{1.3 \rm mm}\simeq400$ mJy.

The emission at 1.3 mm from the outer regions of protoplanetary disks is typically optically thin. Therefore, this flux is commonly used to estimate the dust mass  $M_{\rm dust}$ of the disk \citep[e.g.,][]{Andrews2011}. To convert flux into dust mass, assumptions on disk opacity and dust temperature $T_{\rm dust}$ must be taken. This means that the fluxes shown in Fig.\,\ref{Outer_disk_contrast}c do not necessarily reflect the dust mass of the targets and we cannot firmly conclude that our GII are less massive than the GI. In fact, in a scenario where GII are flat disks and GI are flared disks, the $T_{\rm dust}$ of GII can be smaller since a smaller disk height results in a lower efficiency to heat the disk interior. We note that the estimate on $M_{\rm dust}$ only scales as $T_{\rm dust}^{-1}$. This means that to account for the above-mentioned factor of 7 difference in flux between the GI and GII in our sample, the $T_{\rm dust}$ of GII should be approximately as many times lower as that of GI. Also, as mentioned in the Introduction the dichotomy between flared and flat disks is no longer obvious, and the different millimeter fluxes of the two groups may actually support the view that GII are more compact than GI (and thus have higher disk opacities) and/or that they are less massive in dust than (most of) the GII, as discussed in Sect.\,\ref{Discussion_types}.    



\begin{figure*}
 \centering
 \includegraphics[width=9cm]{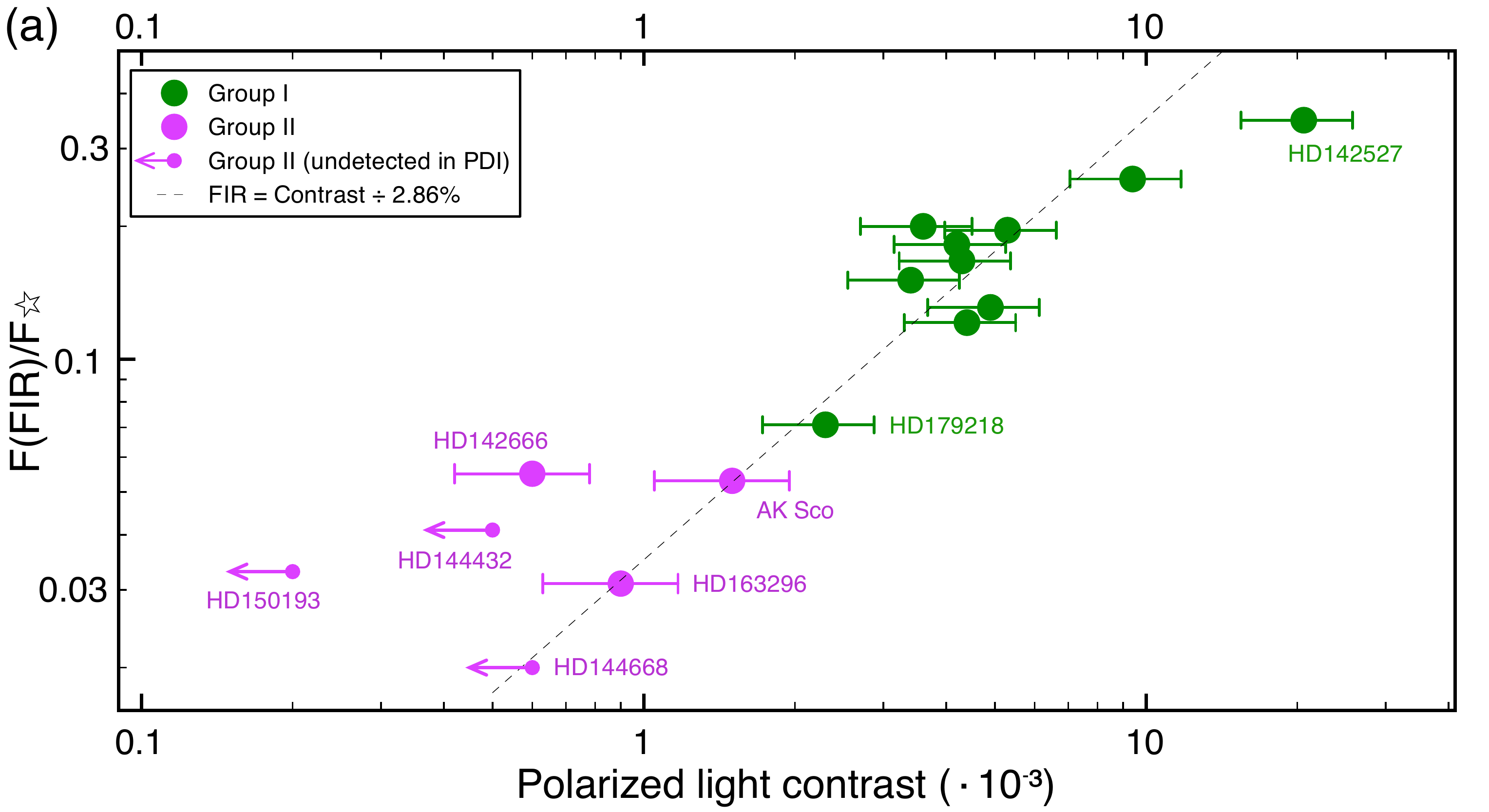} 
   \includegraphics[width=9cm]{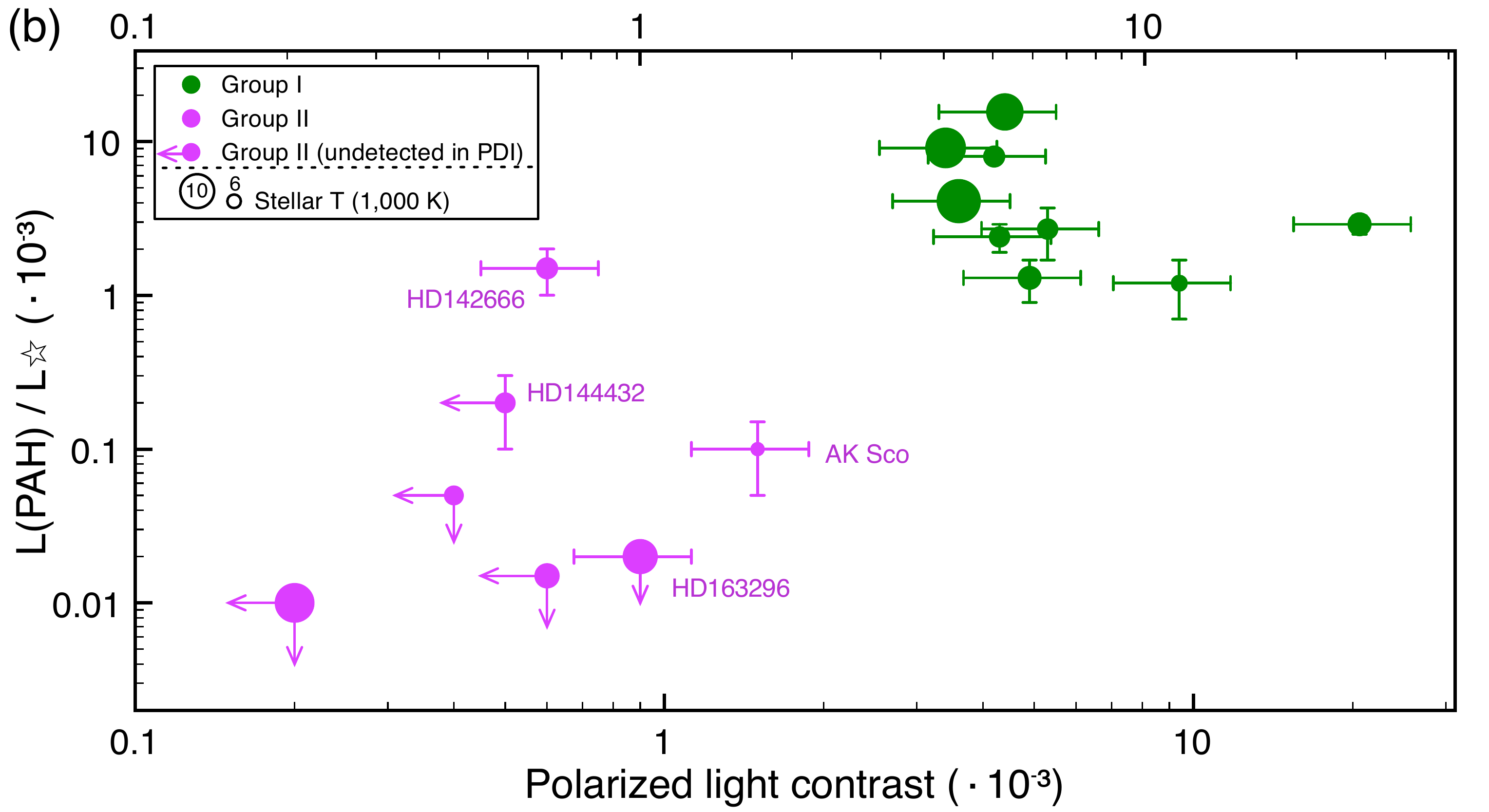} 
  \includegraphics[width=9cm]{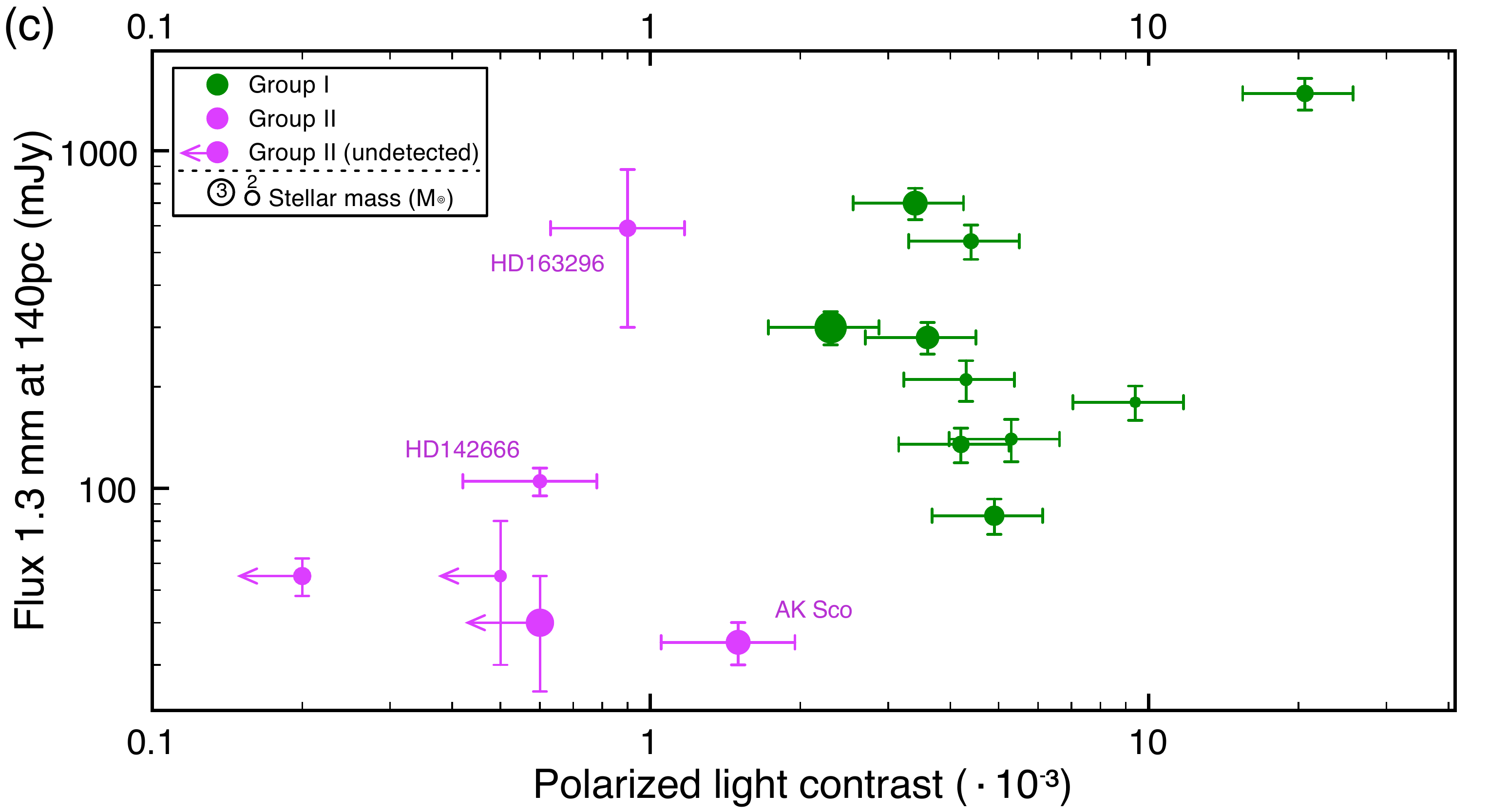} 
  \includegraphics[width=9cm]{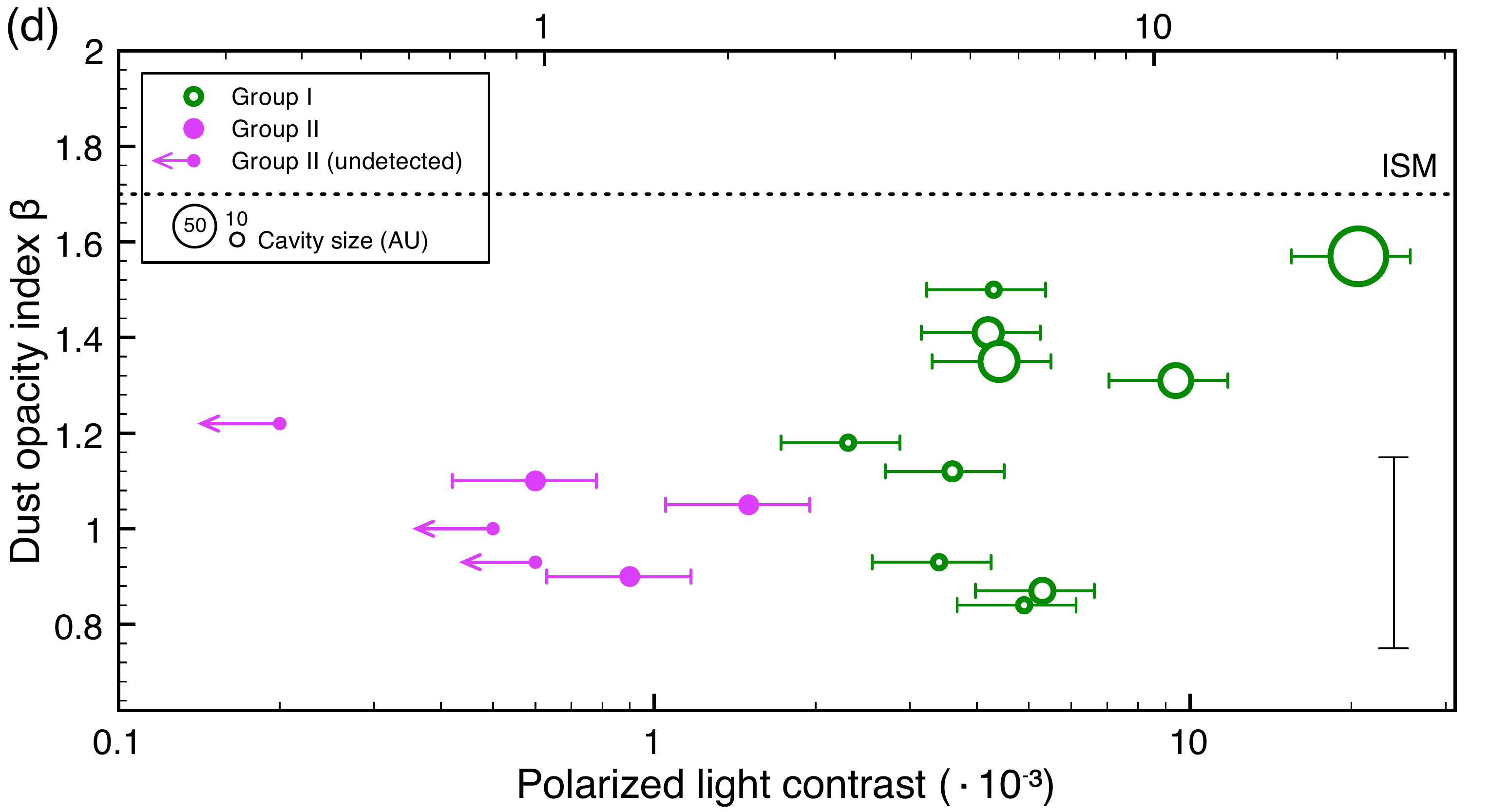} 
  \caption{Outer disk properties of the sample compared with the contrast. \textbf{(a)}: Far-IR excess normalized to the stellar flux. \textbf{(b)}: PAH luminosity normalized to the stellar flux. The symbol size reflects the stellar temperature with dynamic range between 10470 K and 6450 K. \textbf{(c)}: Flux at 1.3mm normalized at a distance of 140pc. The error on the y-axis reflects the uncertainty on the distance (typically smaller for GAIA measurements). The symbol size reflects the stellar mass with dynamic range between 1.6 M$_{\odot}$ and 3.2 M$_{\odot}$. \textbf{(d)}: Dust opacity index $\beta_{\rm mm}$. The symbol size reflects the cavity size from the PDI images. Typical errors for this source type  are indicated by the vertical bar to the right.} 
          \label{Outer_disk_contrast}
  \end{figure*}

\subsubsection{Millimeter index} \label{Sect_Millimeter_index}

In Fig.\,\ref{Outer_disk_contrast}d we show the dust opacity index $\beta_{\rm mm}$ compared with the contrast. If the emission is optically thin,
this index is related to the slope of the millimeter SED $\alpha_{\rm mm}$, via $\alpha_{\rm mm} = \beta_{\rm mm}+2$ \citep[e.g.,][]{Beckwith1991}. Millimeter observations of protoplanetary disks \citep[e.g.,][]{Testi2001} have shown that $\alpha_{\rm mm}$ can be significantly smaller than that of the ISM ($\beta_{\rm mm}\approx1.7$ corresponding to $\alpha_{\rm mm}\approx3.7$). This has been interpreted as being due to the process of dust grain growth in the disk midplane \citep[see, e.g.,][]{Natta2007}. 

We collected $\alpha_{\rm mm}$ or $\beta_{\rm mm}$ from previous works\footnote{We did not retrieve the errors for all measurements. We assume that an average uncertainty for these relatively bright sources is $\sim$0.4 \mbox{\citep[see, e.g.,][]{Pinilla2014}}.} (see Appendix \ref{Appendix_sample}) and made them uniform to $\beta_{\rm mm}$, as in Fig.\,\ref{Outer_disk_contrast}d. It turned out that the average $\beta_{\rm mm}$ of the GI in our sample is only marginally larger than that of GII (1.21 against 1.03, with single uncertainties of $\sim$0.4). Similar trends have been found in the past and have been ascribed to the process of grain growth occurring during the transition from GI to GII \citep[e.g.,][]{Acke2004}. However, Fig.\,\ref{Outer_disk_contrast}d also shows that generally objects with large $\beta_{\rm mm}$ are those with large inner cavities. This can be a consequence of a pressure bump at the outer edge of a disk cavity, which can act to deplete the inner disk of millimeter grains and thus result in a higher $\beta_{\rm mm}$ inside the cavity. This effect was modeled and also seen in the possible trend between $\beta_{\rm mm}$ and cavity size of a large sample of disks by \citet{Pinilla2014}. {Alternatively, lower $\beta_{\rm mm}$ values from disks without a large cavity can be explained by the possible existence of optically thick central regions \citep[as shown by, e.g.,][]{Isella2016}. In fact, the resolved $\beta_{\rm mm}$ values of these regions would be $\sim$0 and would contribute to decreasing the unresolved measurement used in this context.} 

In view of this, the global $\beta_{\rm mm}$ is not a good tracer of the global grain growth, but may only reflect the {different morphology of the two groups}. In fact, if we compare the $\beta_{\rm mm}$ of GII with that of GI without a large cavity in scattered light ($R>15$ AU), we obtain similar values (1.03 against 1.07). Therefore, we propose that the different $\beta_{\rm mm}$  seen in larger sample of GI and GII may only be  the consequence of the dust grain differentiation {and/or the absence of an optically thick central region} in gapped disks rather than a real indication of different evolutionary stages (see Sect.\,\ref{Discussion_evolution}).

\subsection{Inner disk properties} \label{Inner_disk_properties}
The spectral properties from the visible to the mid-IR constrain the morphology of the inner disk and of the immediate surrounding of the star. In this section and in Fig.\,\ref{Inner_disk_contrast} we compare some spectral properties with the contrast and the stellar properties.

\subsubsection{Near-IR excess}
{We calculated the near-IR excess from 1.2 $\mu$m to 4.6 $\mu$m of all sources, following the method described by \citet{Pascual2016}.} This excess is shown in Fig.\,\ref{Inner_disk_contrast}a. Three clusters of datapoints stand out from the plot: GII disks with mid to high near-IR flux, half of the GI with low near-IR flux, and the other half with high near-IR flux. In the plot, we also indicate the presence of features on the disk surface to highlight that three of the four GI with high near-IR flux have a double-arm spiral structure. The fourth member of this cluster, HD142527, also shows multiple spirals, but with different opening angles and at larger radii.

The near-IR flux of the GII in the diagram is  intermediate between that of the two clusters of GI (with the exception of HD144668). Therefore, on average the thermal emission of hot dust from GI and GII is to first order comparable, contrary to the far-IR and the millimeter flux. All GII show a relatively high near-IR excess indicating a recurrently large contribution to the SED from hot particles (see Sect.\,\ref{Discussion_inner}).

\subsubsection{Mass accretion}
In Fig.\,\ref{Inner_disk_contrast}b we show the mass accretion rate calculated from the UV excess by \citet{Fairlamb2015} for some of the sources in our sample. The highest rates are found around more massive stars. There is no significant difference between GI and GII; the former group has an average value of $0.78 \cdot 10^{-7}\ {\rm M_{\odot}}$/year (6/8 objects accreting) and the latter of ${1.95 \cdot 10^{-7}}\ {\rm M_{\odot}}$/year (4/6 objects accreting). Furthermore, there is no correlation between the cavity size and the accretion rate.

{This finding is in contrast with the idea that gapped disks may have a lower accretion rate than continuous disks, as shown in the context of T Tau stars by, e.g., \citet{Najita2015} and \citet{Kim2016}. The small sample does not allow us to determine whether this may reflect a different behavior for the accretion rate of Herbig and T Tau stars.}

\subsubsection{CO ro-vibrational lines}
\begin{figure*}
 \centering
\includegraphics[width=9cm]{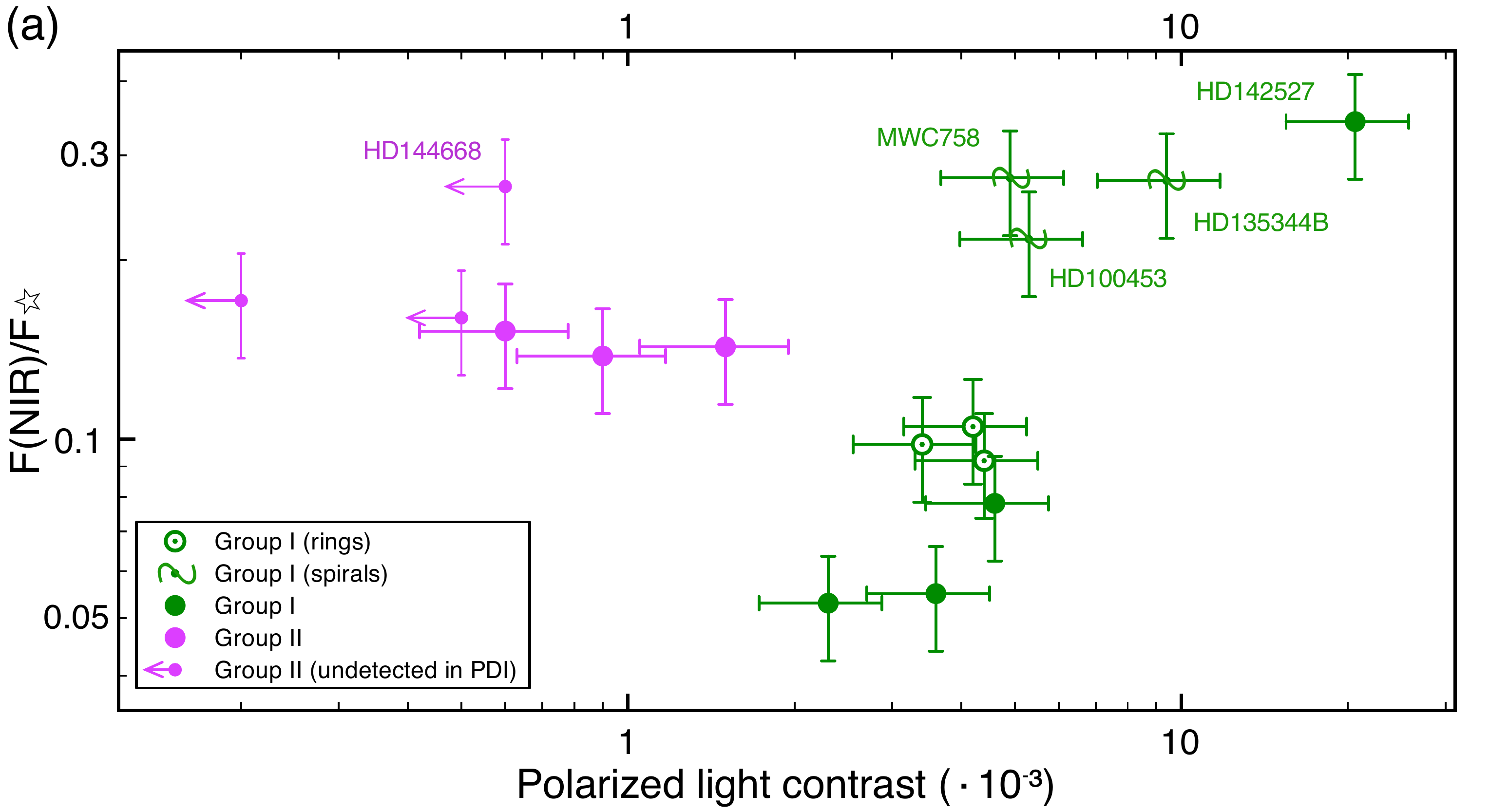} 
 \includegraphics[width=9cm]{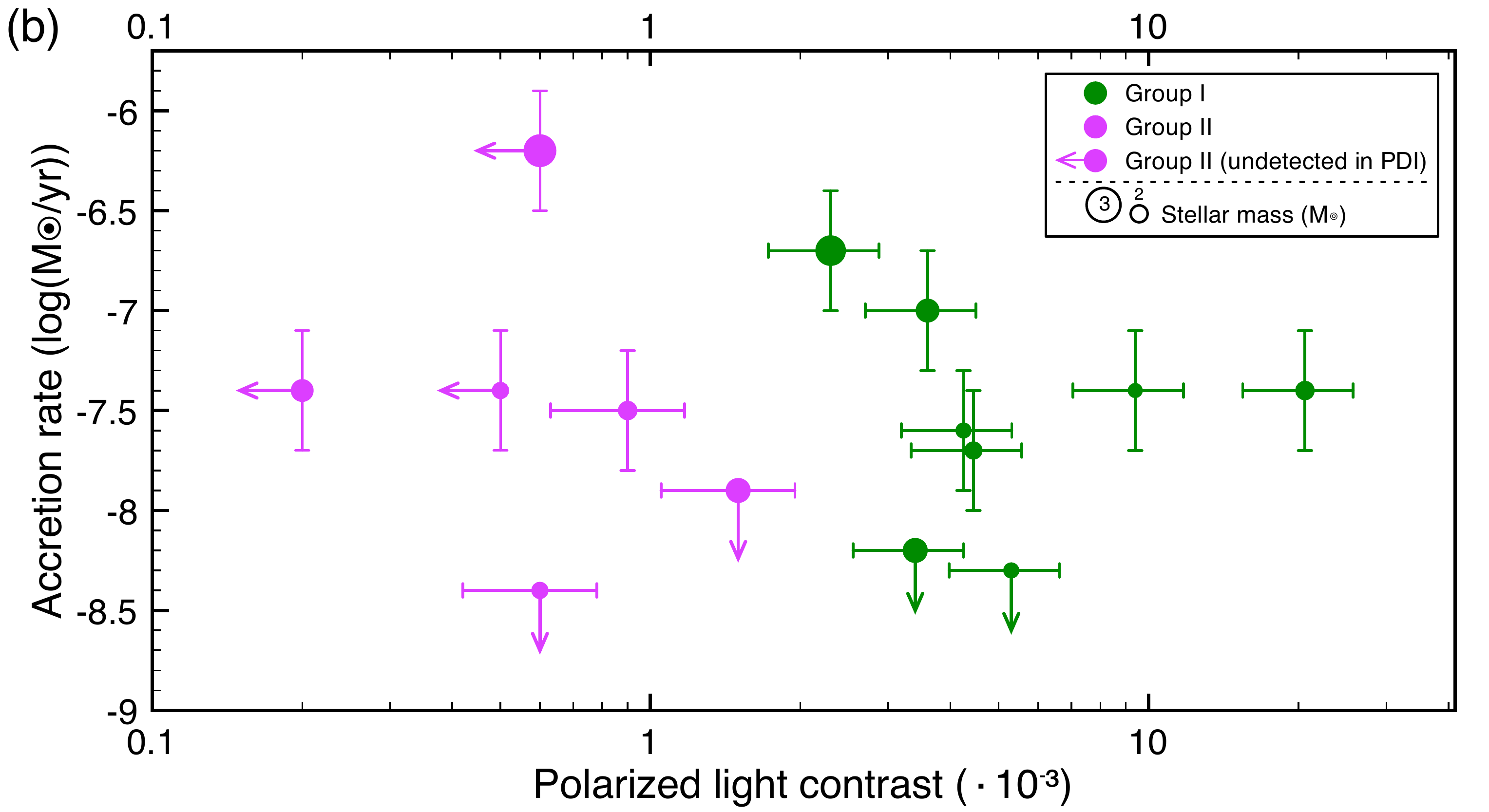} 
 \includegraphics[width=9cm]{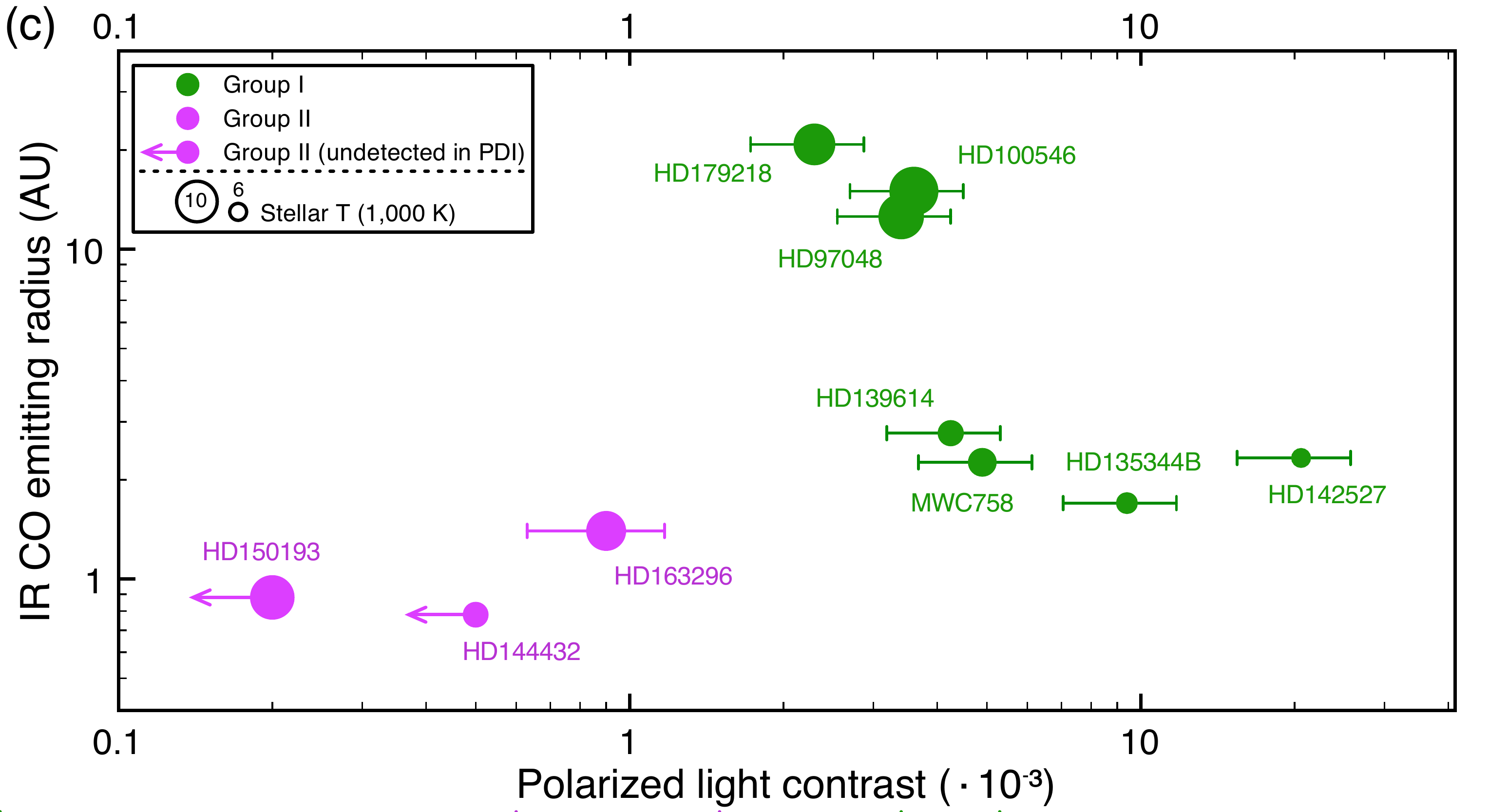} 
 \includegraphics[width=9cm]{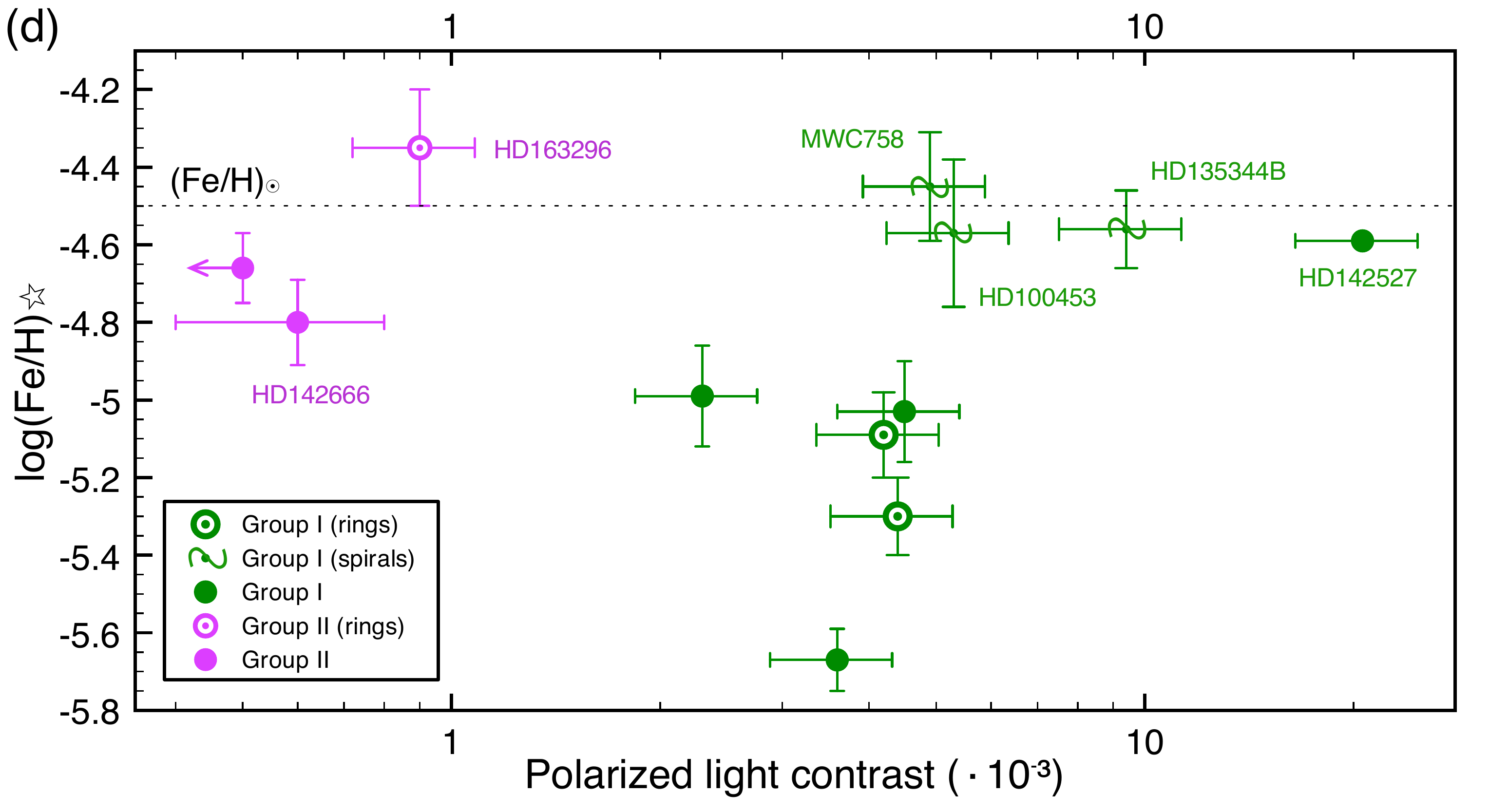}
   \caption{Inner disk properties of the sample compared with the contrast. \textbf{(a)}: Near-IR excess normalized to the stellar flux. \textbf{(b)}: Mass accretion rate. The symbol size reflects the stellar mass with dynamic range between 1.6 M$_{\odot}$ and 3.2 M$_{\odot}$. \textbf{(c)}: Hot CO emitting radius. The symbol size reflects the stellar temperature as in Fig.\,\ref{Outer_disk_contrast}(b). \textbf{(d)}: Stellar photospheric abundance of iron relative to hydrogen. The dashed horizontal line indicates the solar abundance.} 
          \label{Inner_disk_contrast}
  \end{figure*}

CO ro-vibrational lines in the near- and mid-IR trace the hot gas in the very inner disk. From these lines, a characteristic emitting radius for the hot CO can be measured \citep[e.g.,][]{Banzatti2015, Banzatti2016}. These radii for some of our sources are shown in Fig.\,\ref{Inner_disk_contrast}c. Similarly to the PAH strength, this property is linked to the stellar temperature in GI, with the three early stars in our sample showing emitting radii as large as 10 AU or more. On the other hand, the emitting radii of the late stars lie between 1.7 and 2.6 AU.

The three GII shown in the diagram all have slightly smaller CO radii (from 0.8 to 1.4 AU) than all GI, even though the stars are warmer or comparable to the late stars of the GI. 

\subsubsection{Stellar photospheric abundance}
In massive stars, which have long convective mixing times, the photospheric abundance of refractory elements show a correlation with the structure of the inner disk \citep{Kama2015}. This is likely because the depletion of large grains in the inner regions of GI and a few GII disks leads to an increased gas-to-dust ratio in the material accreting onto the star, which may in turn be connected to the trapping of large dust grains by substellar companions. In Fig.\,\ref{Inner_disk_contrast}d we show the stellar photospheric abundance of the iron relative to that of the hydrogen for the stars in our sample. 

Similarly to both Fig.\,\ref{Inner_disk_contrast}a (the near-IR flux) and Fig.\,\ref{Inner_disk_contrast}c (the CO radius), three clusters of sources are visible. The GII and four of the GI show a solar  abundance of iron or slightly lower. The other five GI in the diagram show a significantly depleted abundance. Of particular interest is  that the four GI with a solar abundance of iron are the same four GI with high near-IR excess, and the three of them present in the CO diagram all show small CO radius. Thus, these three diagrams seem to reveal a physical connection between the stellar photospheric abundance of heavy elements, the near-IR excess, and the emitting radius of hot CO. In this regard, the GII of our sample are similar to roughly half of the GI, namely with high near-IR, high [Fe/H], and small CO emitting radius.

\section{Discussion}   \label{Sect_Discussion}
Keeping in mind the small number of objects, the following results on the taxonomy of GI and GII disks are to be considered:  

\begin{enumerate}[(a)]

\item What gives rise to the observed features defining GI and GII is the presence or absence of a disk cavity ($\gtrsim$ few AU large).   

\item Most sources (but not all of them, see HD150193) have polarized contrast scaling with the far-IR excess.

\item Most non-detected GIIs have a stellar companion at 100s AU. The GIs with a companion are the smallest disks in extent.

\item GIIs  typically have weaker millimeter fluxes. However, one GII (HD163296) has the third highest flux in the sample.  

\item If disks with large cavities ($R \gtrsim 30$ AU) are not considered, GIIs have on average the same opacity index as GIs.

\item GIs and GIIs are indistinguishable in terms of mass accretion rate. There is no relation for this rate with the cavity size.

\item GIIs show high near-IR excess, IR CO emission on small radii, and solar photospheric abundance of iron. Four (out of nine) GIs have the same properties (and their outer disks all show spirals), whereas the other GIs have low near-IR excess, CO from larger radii, and depleted abundance of iron.

\end{enumerate}

The implications of these seven major results are discussed in this section with the aim of providing an explanation for the elusiveness of GII disks in scattered light, as well as insight into the nature of the GI-GII dichotomy.
   
\subsection{GI vs GII: gapped vs continuous disks} \label{Discussion_gapped}
Figure \ref{Contrast_color} confirms what was proposed in previous works \citep[e.g.,][]{Currie2010, Maaskant2013, Menu2015}, namely that GI sources are gapped disks whereas GII sources are continuous disks. In fact, all GI disks in our sample show the presence of a \textit{large} cavity ($R\gtrsim 5$ AU) and most of them from either millimeter continuum imaging or \mbox{near-IR} PDI. In two cases to date (HD179218 and HD139614), the cavity has only been   claimed from SED or spectral line fitting \citep{Fedele2008, Carmona2016}. On the other hand, only \textit{small} cavities ($R\lesssim 1$ AU) have been claimed for the GII disks in our sample \citep[e.g.,][see Sect.\,\ref{Sect_Sample}]{Menu2015}. For simplicity, we refer to disks with large cavities as gapped disks and to disks with small cavities as continuous disks.

Once the connection GI$\equiv$gapped and GII$\equiv$continuous disks is established, it is clear that the study of the observational properties differing in the two groups must be revised in the context of the new dichotomy. For example, treating GI as gapped disks may also explain the incongruous PAH brightness in GIs and GIIs raised by \citet{Dullemond2007}. In fact, their models show that if GIIs are the result of dust sedimentation occurring in GIs, their PAH brightness should be enhanced because of the reduced opacity, and thus increased UV radiation, in the environment where the PAH luminosity originates. However, the observations show the opposite trend \citep[and Fig.\,\ref{Outer_disk_contrast}b of this work]{Acke2010}. We speculate that, if instead the GIs are gapped disks and GIIs are not, an increased amount of UV-exposed PAH molecules is to be expected in GIs, reconciling theory and observations. This hypothesis has an intriguing consequence related to the weak detection of PAH emission from three GIIs only. In fact, HD142666 and HD144432 are the GIIs in our sample where a small disk cavity has been claimed \citep{Chen2012, Menu2015}. The third, AK Sco, is composed of binary stars separated by 0.16 AU and an intrinsically larger inner cavity is thus to be expected. Therefore, our dataset suggests that the PAH emission may be intimately related to the presence of an inner cavity and the reason is an increased UV radiation in disks with lower optical depth at small radii.

More generally, it is not obvious whether some disk properties (see Table \ref{Literature}) are ($i$) the result of the geometry provided by the presence/absence of a cavity, or that ($ii$) they trace the disk conditions that allow or not the formation of a cavity. For  example, is the lower scattered light of GII disks due to self-shadowing in continuous disks ($i$) or does it reflect a different geometry for disks (maybe smaller or less massive) that cannot open large cavities ($ii$)? The former explanation points toward GI and GII being different evolutionary \textit{stages}, whereas the latter points toward them being different evolutionary \textit{tracks} in the disk lifetime. These two scenarios are discussed in Sect.\,\ref{Discussion_evolution}. Here we stress that the answer to this question is intimately related to the origin of disk cavities, which is in turn a longstanding debate in the disk community. Interactions with orbiting companions \citep{Rice2003}, photoevaporation \citep{Alexander2006}, and dust grain growth \citep{Dullemond2005} are only some of the proposed explanations for the disk cavities. With specific focus on the objects of our sample, the literature indicates an increasing consensus on the interaction with (forming) planets as the most probable cause. Also, two GIs in our sample (HD100546 and HD169142) may have a detected  substellar companion within the cavity \citep{Brittain2014, Reggiani2014, Biller2014}.

In this scenario,  the different cavity sizes for $\mu$m- and mm-sized dust grains are also fundamental constraints as they may indicate a pressure bump at the outer edge of the cavity that filters grains with different sizes \citep[e.g.,][]{Pinilla2012, Garufi2013}. In our sample, these differences are varied. Two extreme cases are HD97048 and MWC758, where millimeter imaging indicates cavities as large as 40 AU and 55 AU \citep{vanderPlas2016, Andrews2011}, but PDI images trace $\mu$m-sized grains at least as close to the star as $\sim$15 AU \citep{Ginski2016, Benisty2015}. We defer an in-depth analysis of these differential cavity sizes to a specific work on gapped disks, and stress that the cavity sizes shown in Fig.\,\ref{Contrast_color} are not from a homogenous observational techniques and should  thus be taken for qualitative consideration only.

\subsection{GI and GII: evolutionary stages or evolutionary tracks?} \label{Discussion_evolution}
As noted in the introduction, GI disks were initially thought to be precursors of GII disks in the framework of the vertical settling of dust grains with time \citep[e.g.,][]{Dullemond2004}. However, the notion that all GIs are gapped disks discredits this scenario. 

One of the observational pieces of evidence supporting the evolution from GI to GII was that GIs have on average smaller grains than GIIs \citep{Acke2004}. In Sect.\,\ref{Sect_Millimeter_index} we confirm this trend (see Fig.\,\ref{Outer_disk_contrast}d), but also show that this is most likely entirely due to the presence of a cavity, since comparing the dust opacity index $\beta_{\rm mm}$ of GII only to GI with small cavities results in comparable $\beta_{\rm mm}$ values. This result can be explained by the presence of a pressure bump at the inner edge of gapped disks, which filters large grains and thus produces a large disk region populated by smaller grains only \citep{Pinilla2014}. {Alternatively, the discrepancy can be due to the presence of optically thick central regions that contribute to lowering the $\beta_{\rm mm}$ of sources without a central cavity.} In other words, GIs may show {smaller $\beta_{\rm mm}$ values than GIIs because they have different disk morphologies} and not because they are at an earlier stage of global dust grain growth.   

If GIIs are not evolved GIs, it can  even be hypothesized that the disk evolution proceeds instead from GII to GI in a scenario where the formation of an increasingly large cavity acts to illuminate the outer disk. However, the low millimeter fluxes (and thus dust masses, see Fig.\,\ref{Outer_disk_contrast}c) and the small radial extent (where detected, see Fig.\,\ref{Contrast_stellar}b) of the GIIs in our sample (except HD163296) rules out this possibility. If the evolution in both directions is excluded, then GIs and GIIs are likely different evolutionary tracks, as proposed by \citet{Currie2010} and \citet{Maaskant2013}. Nonetheless, the scenario where HD163296 is a precursor of the GI cannot be ruled out and is discussed in Sect.\,\ref{Discussion_types}.

Owing to the large uncertainties on stellar ages, there has not been any conclusive evidence that GIIs are older than GIs. The stellar ages of our sample vary enormously from work to work and thus we do not draw on this property. However, we note that the range of ages of GIIs is between 2 Myr and 6 Myr, whereas that of GIs is between 1 Myr and 15 Myr with almost half of them  aged $\geq$10 Myr. Thus, the possibility that (some) GI disks may be longer lasting structures should be cautiously considered \citep[see also][]{Kama2015}. This longevity could be  explained by the possible presence of planetary bodies within the disk cavity that prevent the rapid accretion of outer material.

\subsection{Different types of GII} \label{Discussion_types}
The taxonomic analysis of Sect.\,\ref{Sect_Taxonomy} reveals that most properties of GII disks are comparable within the group, with dispersions lower than one order of magnitude for the entire sample. However, the following properties strongly vary within the group, and may indicate the need for a subclassification: 

\begin{itemize}

\item Disk extent. In scattered light, one source is large (HD163296), two are smaller (AK Sco and HD142666), while four are not detected. To date, the only resolved observations in the millimeter were obtained for HD163296.

\item Millimeter flux. One source (HD163296) is very bright, whereas five are faint. Within the latter category, HD142666 has a higher flux that is comparable to the fainter GIs.

\item PAH emission. In one case (HD142666) the emission is comparable to the GIs around relatively late-type stars, whereas HD144432 and AK Sco show significantly lower emission and four sources remain undetected.

\item Mass accretion rate. The sources span more than two orders of magnitude, from the very high rate of HD144668 (which is the most massive star in the sample) to the non-detections of AK Sco and HD142666.  

\end{itemize}

An evident dichotomy arises between HD163296 and the rest of the GIIs. Keeping in mind the uncertainty of converting millimeter fluxes into dust masses, HD163296 is likely more massive in dust than the other objects. Furthermore, it is known to have a gaseous disk that is twice as large as the dusty disk \citep{deGregorio-Monsalvo2013} and to host a prominent jet \citep[e.g.,][]{Ellerbroek2014}. Scattered light images trace small dust grains as  far out as hundreds of AU \citep{Grady2000}. All in all, the only properties that distinguish HD163296 from the other GI are those connected to the illumination of the outer disk (scattered light, far-IR, and PAH), which points toward a self-shadowed disk. Considering the absence of a large disk cavity \citep{deGregorio-Monsalvo2013} and the presence of a strong jet, it is possible that HD163296 is a precursor of the classical GI. The existence of rings in both scattered light and millimeter continuum images \citep[this work and][]{Isella2016} reinforces the analogy with the GIs, which often show these features \citep[e.g.,][]{Ginski2016}.   

On the other hand, the detection in scattered light of the disk around HD142666 and AK Sco (see Fig.\,\ref{Imagery}) may constrain their outer edge to a few tens of AU. Millimeter imaging of AK Sco detects signal on a slightly larger scale, i.e.,\ up to $\sim$100 AU \citep{Czekala2015}. Conversely, our outer edge for the disk of HD142666 is consistent with the cold CO distribution \citep[traced out to $\sim$60 AU,][]{Dent2005}. Their millimeter fluxes are respectively low ($\sim$105 mJy, if scaled at 140 pc) and very low ($\sim$35 mJy). Even though it is not detected in scattered light, HD144432 may be a similar object, having CO traced as out as $\sim$45 AU \citep{Dent2005} and showing PAH emission. As commented in Sect.\,\ref{Discussion_gapped}, the PAH detection is a possible consequence of their small cavities at sub-AU scale. All this may suggest that these disks are slightly smaller counterparts of GIs, which were unable to create a large disk cavity. 

Finally, the disk of HD145263 may be undergoing the final stages of disk dissipation \citep{Honda2004}. The other three non-detected GIIs all have a stellar companion at projected distances of approximately 100 AU. They show very low millimeter fluxes and two of them (HD144668 and HD150193) show no PAH emission. In the case of HD150193, we can infer that the non-detection of the disk in scattered light \citep{Garufi2014b} is inconsistent with the amount of far-IR excess (Fig.\,\ref{Outer_disk_contrast}a) and with the mid-IR [30/13.5] ratio (Fig.\,\ref{Contrast_color}). These notions may suggest that the entire mid- to far-IR excess originates at disk radii smaller than the inner working angle of the PDI observations, namely $\sim$15 AU, which is in agreement with works based on SED fitting \citep{Dominik2003}, and possibly with the non-detection of cold CO \citep{Dent2005}. For HD144668, we cannot infer the same inconsistency because the upper limit on the disk detection (from this work) is higher. However, \citet{Preibisch2006} showed that the mid-IR emission from this disk is confined within 2.5 AU from the star, which is significantly less than for typical disks around Herbig Ae/Be stars. 

All the above considerations seem to indicate the existence of a family of GII disks (like HD150193 and HD144668) with compact disks, having an outer radius on the order of 10 AU or only slightly more, and dust (and possibly gas) masses significantly lower than the GI disks. The most straightforward method for corroborating or rejecting this hypothesis is future millimeter continuum imaging by ALMA. In any case, a relatively small disk is to be expected in HD150193, HD144432, and HD144668 because of the presence of stellar companions. It is typically assumed that circumprimary disks are truncated at roughly 1/3 of the distance to the companion \citep{Artymowicz1994}, meaning that these three disks may still be as large as 30-40 AU, unless the respective companion is in a very eccentric orbit and currently at  aphelion. 

Even though the diverse disk geometries proposed in this section cannot be currently confirmed by the available datasets, it is clear that referring to a unique class of GII objects is too simplistic. At least two (potentially three) types of GIIs likely exist, with self-shadowed and relatively small (and less massive) disks being indistinguishable from their SED. Given the small sample of this work, we cannot conclude whether HD163296 and HD142666 host peculiar disks or whether they are actually part of two large families of disks that are both currently labeled as GII.   

\begin{figure*}
 \centering
\includegraphics[width=18cm]{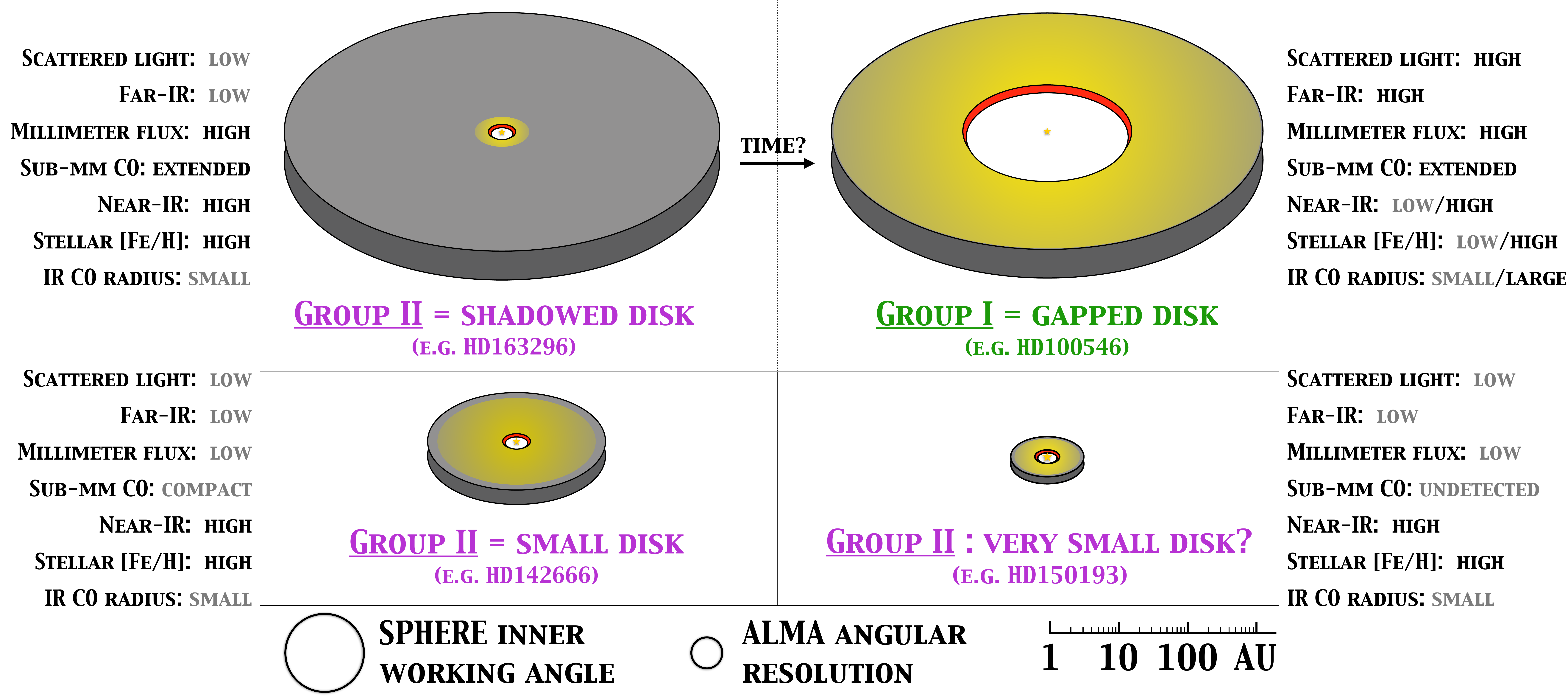} 
   \caption{Summary of the properties of the sources analyzed in this work. The proposed disk geometries are shown in logarithmic scale. The SPHERE inner working angle is imposed by the angular resolution of observations in the near-IR ($\sim$10 AU for sources at $\sim$150 pc). The ALMA angular resolution of $\sim$3 AU is achieved with the longest possible baselines, which should be used to resolve potentially very small disks.} 
          \label{Sketch}
  \end{figure*}

\subsection{ Environment of the innermost disk regions} \label{Discussion_inner}
Since the disks of all GIs are depleted in dust within at least 10 AU of the central star,    different optical/near-IR properties may be expected from the GII disks. In Sect.\,\ref{Inner_disk_properties}, we show the existence of three clusters of objects in terms of near-IR excess, emitting radius of CO, and stellar photospheric abundance of iron. To a large extent, the objects in the three clusters are the same and the connection between the properties is real. 
In particular, the near-IR excess is $\gtrsim10\%$ of the stellar flux for all GIIs in our sample (Fig.\,\ref{Inner_disk_contrast}a) and for four (out of nine) GIs: MWC758, HD100453, HD135344B, and HD142527. The outer disk of these stars all show spiral-like features \citep[e.g.,][]{Benisty2015, Benisty2016, Garufi2013, Avenhaus2014a}. All  four of these sources are also relatively late-type stars (see Fig.\,\ref{Contrast_stellar}a). However, other late-type stars in the sample have low near-IR excess, whereas some early-type stars have high near-IR excess. Thus, the dichotomy cannot be explained by the stellar temperature alone.

Why some Herbig Ae/Be stars have very high near-IR excess is a longstanding debate. Hydrostatic disk models typically fail to reproduce it, indicating that the emission is partly due to material uplifted from the disk by a wind \citep{Bans2012} or that the inner disk is composed of refractory elements that are present at smaller radii than the sublimation radius for silicates. In particular, customized works have shown that the latter case could be the explanation for the near-IR flux of the GII HD163296 and HD144668 \citep{Benisty2010a, Benisty2011}. In  the case of uplifted and of inner material, it is not possible to infer from the near-IR excess alone whether this material has the sufficient optical depth to shadow the outer disk region. In any case, it must be noted that the outer disk of three-quarters of the GI with high near-IR flux shows signs of shadows by an inclined inner disk \citep{Marino2015, Stolker2016, Benisty2016}. Speculatively, it may even be possible to connect the presence of these shadows with that of spirals. In fact, both hydrodynamical simulations \citep{Montesinos2016} and scattered light observations \citep{Wagner2015, Benisty2016} show a possible connection between these features, due to the reduced pressure in correspondence of the shadows that can excite spiral arms.   

The physical link between the near-IR flux, the radius of IR CO emission, and the stellar photospheric abundance of iron is not straightforward, and will be discussed in depth in a dedicated work that is in preparation. The measurements included in this work show that whatever process is determining the dust and CO gas emission from the inner disk also has an impact on the stellar photosphere. \citet{Kama2015} proposed that the depleted abundance of refractory elements in the stellar photosphere of GI can be connected to the increased gas-to-dust ratio of the material flowing within a cavity because of the trapping of small dust grains by substellar companions. Following this thinking, the decreased abundance of refractory elements of the inflowing material may result in a reduced near-IR excess, due to the lower sublimation temperature of silicates compared to refractory elements. The distribution of CO in the inner few AU may be also modified, either by the removal of volumes of gas or by a changed interplay with the local dust, or both. A decrease in small dust grains in the inner disk would in fact damp the IR pumping of the surrounding CO and, if the column density of CO gas is reduced, would facilitate the excitation of CO by UV pumping at larger radii \citep[as in HD179218, HD100546, and HD97048, see Fig.\,\ref{Inner_disk_contrast}c and][]{vanderPlas2015}.

\section{Summary and conclusions} \label{Sect_Conclusions}
Since the beginning of the century, the most recognizable evolutionary track of protoplanetary disks around Herbig Ae/Be stars has been thought to be the dust settling that leads  flared disks (Group I)  to evolve into  flat disks (Group II)  \citep{Meeus2001}. In this work, we analyze VLT/NACO \mbox{near-IR} scattered light images of six GIIs with the aim of complementing the available sample of GIs. Even though the observations were carried out in suboptimal conditions, we detect a disk around half of the sources. In particular:

\begin{enumerate}

\item The brightness distribution in the disk around HD163296 is spatially consistent with that by \citet{Garufi2014b}, indicating the persistency of a ring-like structure located slightly inside the CO snowline \citep{Qi2015, Guidi2016}.

\item A relatively small disk ($\sim$60-70 AU) is retrieved around \mbox{AK Sco} and HD142666. The signal from these disks can be traced at least {down to} $\sim$15 AU.

\end{enumerate}

We investigate the different nature of GI and GII disks by means of a taxonomic analysis of 17 sources (10 GIs and 7 GIIs) observed in polarized near-IR light and with stellar/disk properties available from the literature. This sample represents more than a half of all the polarimetric images of protoplanetary disks currently available. With \textit{specific regard} to the analyzed sample, the main results are the following:

\begin{enumerate}

\setcounter{enumi}{2}

\item All GI disks have a cavity larger than $\sim$5 AU, while no GII disk has a cavity larger than $\sim$1 AU. 

\item The amount of far-IR excess and of near-IR scattered light correlates. One significant exception is HD150193.

\item The millimeter flux, to the first order tracing the dust mass, is systematically lower in GIIs than in GIs (by a factor of 6-7). One significant exception is HD163296, which is also the only GII with available resolved observations in the millimeter.

\item The disks around GII objects with a stellar companion at hundreds of AU are not detected in scattered light, while those around GI with similar companions are the smallest in extent.

\item The different dust opacity index $\beta_{\rm mm}$, tracing the grain sizes of GIs and GIIs is likely due to the depletion of large grains within the cavity. It does not necessarily reflect a more advanced stage of global dust grain growth for GIIs. 

\item Keeping in mind the uncertainties on stellar ages, we find no GII older than 6 Myr, but half of the GIs are older than 10 Myr.

\item The PAH luminosity, tracing the volume of gas exposed to UV radiation, is high in all GIs while it is very low in four out of seven GIIs. The peculiarity of the three GIIs with high PAH (HD142666, HD144432, and AK Sco) is that they  host a small-scale cavity. Thus, an analogy between the PAH brightness and the presence of a cavity emerges. This could solve the long-standing inconsistency for the PAH between theory and observations. 

\item We find a clear link between the amount of near-IR excess, the stellar photospheric abundance of iron, and the emitting radius of CO gas as probed in the IR. All GIIs and  half of the GIs show respectively high, high, and small values, while the other half of the GIs show low, low, and large values.


\end{enumerate}   

Point 3\ in the above list indicates that the dichotomy of SEDs shown by Group I and Group II is due to the presence or absence of a large inner cavity, and thus to the different illumination that the outer disk is subject to. Therefore, the evolution from Group I (flared disks) to Group II (flat disks) as a result  of dust settling must be revised. In fact, there is no property supporting this evolutionary track (see Points 7\ and 8).

We also propose that the dichotomy between the millimeter-bright GII HD163296 and the other millimeter-faint GIIs (\mbox{Point 5}) is indicative of very different disk geometries. Some GII disks may be \textit{smaller} version{s} of the GI disks ($<$100 AU in extent and up to one  order of magnitude less massive in dust) that are unable to form large cavities. The disks of HD142666 and AK Sco (Point 2) are the prototypes of these relatively small structures. Our analysis also suggests the existence of very small disks, with $R\sim10$ AU, which would be undetectable in scattered light and would thus explain the outlier of Point 4. These disks are possibly subject to truncation by stellar companions (\mbox{Point 6}) and can be currently only imaged by ALMA. On the other hand, HD163296 shows the same properties as the GI, with the exception of those properties that are related to the disk illumination (far-IR, scattered light, PAH, see Point 9). It can  therefore be a primordial version of the GI, with a prominent jet and a continuous disk that efficiently shadows its outer regions \citep[Point 1\ and][]{Garufi2014b}. In Fig.\,\ref{Sketch}, we show a sketch summarizing the proposed disk geometries.

Finally, the dichotomy between sources with high and low near-IR excess (Point 10) may provide new insight into the process of planet formation within the disk cavities. We hypothesize that the amount of near-IR flux is related to the abundance of refractory elements in the inflowing material and that this also has an imprint on the stellar photospheric abundance of elements. A possible connection between the morphology of the inner and outer disk is also proposed, with those sources with high near-IR excess also showing shadows and spirals in scattered light.

Follow-up studies are needed to understand whether the conclusions of this paper also apply to a larger sample of protoplanetary disks. It is of particular importance to extend the study to lower mass stars (the T Tauri stars) and to sources with evidence of primordial jets in order to obtain deeper insight into evolutionary tracks and evolutionary stages of disks throughout the planet formation. Finally, ALMA observations of those disks that remain undetected in scattered light are fundamental in order to disentangle their morphology and to provide a view of the variety of protoplanetary disks that is less biased toward particularly bright and extended objects.

\begin{acknowledgements}
We thank L.B.F.M.\,Waters, I.\,Kamp, A.\,Carmona, and L.\,Klarmann for clarifying discussions. We acknowledge the referee for the interesting comments. We are grateful to the SPHERE consortium, and in particular to C.\,Ginski and J.\,de Boer, for useful discussions and for making the SPHERE GTO data available for the paper. Part of this work has been carried out within the framework of the National Centre for Competence in Research PlanetS supported by the Swiss National Science Foundation. S.P.Q.\ and H.M.S.\ acknowledge the financial support of the SNSF. H.C.\ and G.M.\ acknowledge support from the Spanish Ministerio de Economia y Competitividad under grant AYA 2014-55840-P. G.M.\ is funded by the Spanish grant RyC-2011-07920. The authors acknowledge the staff at VLT for their excellent support during the observations. This research has made use of the SIMBAD database, operated at CDS, Strasbourg, France.
\end{acknowledgements}

\bibliographystyle{aa} 
\bibliography{Reference.bib} 


\begin{appendix}

\section{The sample} \label{Appendix_sample}
The sample studied in this work consists of 17 B, A, and F stars (10 GI and 7 GII) that were  observed in near-IR PDI with either VLT/NACO or VLT/SPHERE from 2012 to 2016. Properties and references of all sources are shown in Table \ref{Table_sample}.1.

\begin{sidewaystable*} \label{Table_sample}
\centering
\begin{tabular}{c|cccccccccccccccc}
\hline
\noalign{\smallskip}
HD & Alternative & Group & Contrast & $d$ & $T_{\rm eff}$ & $M_*$ & [30/13.5] & F$_{\rm NIR}$/F$_*$ & F$_{\rm FIR}$/F$_*$ & F$_{\rm 1.3mm}$ & $\beta_{\rm mm}$ & L$_{\rm PAH}$/L$_*$ & log($\dot{M}$) & $R_{\rm CO}$ & $R_{\rm gap}$ & log(Fe/H)  \\
name & name & & ($\cdot 10^{-3} $) & (pc) & (K) & ($\rm M_{\odot}$) & & (\%) & (\%) & (mJy) & & ($\cdot 10^{-3}$) & ($\rm M_{\odot}$/yr) & (AU) & (AU) & \\
\noalign{\smallskip}      
            \hline
            \hline
            \noalign{\smallskip}
34282 &  & I & 4.4 $\pm$ 0.8 & 325 & 9,500 & 1.9 & 10.0 & 9.2 & 12.1 & 100 & 1.05 & 15.6 & -7.7 & & 70$^p$ & -5.30 \\
36112 & MWC758 & I & 4.9 $\pm$ 1.6 & 151 & 7,750 & 2.2 & 4.1 & 27.5 & 13.1 & 72 & 0.85 & 1.3 & & 2.26 & 55$^m$ & -4.45 \\
97048 &  & I & 3.4 $\pm$ 1.2 & 179 & 10,000 & 2.5 & 5.9 & 9.8 & 15.1 & 452 & 0.93 & 9.0 & <-8.2 & 12.57 & 40$^m$ & \\
100453 &  & I & 5.3 $\pm$ 1.7 & 102 & 7,400 & 1.7 & 5.2 & 21.7 & 19.6 & 265 & 0.87 & 4.7 & <-8.3 & & 18$^p$ & -4.57 \\
100546 &  & I & 3.6 $\pm$ 1.4 & 109 & 10,470 & 2.4 & 3.5 & 5.4 & 20.5 & 465 & 1.12 & 4.1 & -7.0 & 15.0 & 14$^p$ & -5.67 \\
135344B & SAO206462 & I & 9.4 $\pm$ 2.4 & 156 & 6,810 & 1.6 & 10.9 & 27.2 & 25.6 & 142 & 1.31 & 1.8 & -7.4 & 1.70 & 45$^m$ & -4.56 \\
139614 &  & I & 4.3 $\pm$ 1.4 & 131 & 7,400 & 1.7 & 4.2 & 7.8 & 16.7 & 242 & 1.50 & 2.5 & -7.6 & 2.77 & 6$^s$ & -5.03 \\
142527 &  & I & 20.6 $\pm$ 2.2 & 156 & 6,550 & 2.0 & 5.0 & 34.2 & 34.8 & 1190 & 1.57 & 2.9 & -7.4 & 2.33 & 140$^p$ & -4.59 \\
142666 &  & II & 0.6 $\pm$ 0.3 & 150 & 7,500 & 1.8 & 1.5 & 15.2 & 5.5 & 99 & 1.10 & 1.9 & <-8.4 &  & - & -4.80 \\
144432 &  & II & $<$ 0.5 & 160 & 7,345 & 1.8 & 1.8 & 16.0 & 4.1 & 44 & 1.00 & 0.3 & -7.4 & 0.78 & - & -4.66 \\
144668 & HR 5999 & II & $<$ 0.6 & 163 & 7,925 & 3.2 & 1.0 & 26.6 & 1.9 & 34 & 0.93 & - & -6.2 & & - &  \\
145263 &  & II & $<$ 0.4 & 135 & 7,200 & 2.0 & 2.0 & & & & & - & & & - & \\
150193 &  & II & $<$ 0.2 & 145 & 9,870 & 2.3 & 1.4 & 17.1 & 3.3 & 45 & 1.22 & - & -7.4 & 0.88 & - & \\
152404 & AK Sco & II & 1.5 $\pm$ 0.4 & 144 & 6,450 & 2.5 & 3.3 & 14.3 & 5.3 & 33 & 1.05 & 0.1 & <-7.9 & & - & \\
163296 &  & II & 0.9 $\pm$ 0.3 & 122 & 9,250 & 2.0 & 2.0 & 13.8 & 3.1 & 780 & 0.90 & - & -7.5 & 1.40 & - & -4.35 \\
169142 &  & I & 4.2 $\pm$ 0.3 & 117 & 7,500 & 2.0 & 7.8 & 10.2 & 18.2 & 197 & 1.40 & 6.1 & & & 16$^p$ & -5.09 \\
179218 &  & I & 2.3 $\pm$ 0.5 & 290 & 9,500 & 3.0 & 2.4 & 5.4 & 7.2 & 71 & 1.18 & & -6.7 & 20.80 & 15$^s$ & -4.99 \\

\hline
 \noalign{\bigskip}
 \noalign{\bigskip}
\end{tabular}

\caption{Properties of the sample. Columns are: name of the source; alternative name commonly used in the literature; Group; polarized-to-stellar light contrast, calculated in this paper from the works listed below; distance, from \citet{Gaia2016}, except 144432 and 163296 \citep{Esa1997} and 144668 \citep{vanLeeuwen2007}; effective temperature of the main star, from \citet{Pascual2016}, except 34282 from \citet{Fairlamb2015} and 144432, 145263, 152404 from \citet{Acke2010}; stellar mass, ibid.\ except 144432 from \citet{Mueller2011} and 145263 from \citet{Sylvester2000}; $[30\, \mu m/13.5\, \mu m]$ ratio, from \citet{Acke2010}; near-IR and far-IR excess normalized to the stellar flux, {from this work following the method by \citet{Pascual2016}}; flux at 1.3 mm, from the references below; dust opacity index, adapted from \citet{Pascual2016}, except 34282, 144432, 145263, 152404 from \citet{Acke2004} and 36112 from \citet{Pinilla2014}; PAH luminosity normalized to the stellar flux, from \citet{Acke2010}, except 100546 from \citet{Meeus2013}; mass accretion rate, from \citet{Fairlamb2015}; CO emitting radius, from \citet{Banzatti2015} and \citet{Banzatti2016}, except 100546, from \citet{vanderPlas2015}; cavity size from millimeter continuum (m), PDI image (p), or SED fitting (s) as from below; stellar photospheric abundance of iron relative to hydrogen, from \citet{Kama2015}. References for the PDI images used for the contrast and for the cavity size (when this is from the same paper, we omit the repetition): 34282, SPHERE consortium in prep.; 36112, \citet{Benisty2015}, \citet{Andrews2011}; 97048, \citet{Ginski2016}, \citet{vanderPlas2016}; 100453, \citet{Benisty2016}; 100546, \citet{Garufi2016}; 135344B, \citet{Garufi2013}, \citet{Andrews2011}; 139614, SPHERE consortium in prep., \citet{Carmona2016}, 142527, \citet{Avenhaus2014a}; 142666, 144432, 144668, this work; 150193, \citet{Garufi2014b}; 152404, 163296, this work; 169142, \citet{Quanz2013b}, 179218, SPHERE consortium in prep., \citet{Fedele2008}. References for the millimeter fluxes: 34282, \citet{Natta2004}; 36112, \citet{Mannings2000}; 97048, \citet{Henning1994}; 100453, \citet{Meeus2012}; 100546, \citet{Henning1994}; 135344B, \citet{Sylvester1996};  139614, \citet{Sylvester1996}; 142527, \citet{Walker1995}; 142666, \citet{Meeus2012}; 144432, \citet{Walker1995}; 144668, \citet{Meeus2012}; 150193, \citet{Sandell2011}; 152404, \citet{Czekala2015}; 163296, \citet{Mannings1997}; 169142, \citet{Sylvester1996}; 179218, \citet{Mannings2000}.  }
\end{sidewaystable*}

\section{Polarized-to-stellar light contrast} \label{Appendix_contrast}
To evaluate the amount of scattered light from a large dataset, we estimate the contrast of the polarized flux from the $Q_{\phi}$ image with respect to the central star. Specifically, we perform a cut with width equal to the resolution along the direction of the disk major axis (where known). The choice of the major axis is dictated by the minimized impact of the disk inclination, which can significantly alter the amount of scattered light along all the other directions. We multiply the extracted flux $F_{\rm pol}$ by the squared distance $r$ from the star, to compensate for the dilution of the stellar radiation, and average radially. To remove the impact of the disk extent, the average is performed between two specific radii, $r_{\rm in}$ and $r_{\rm out}$, different for each source and set by the disk inner edge and the outermost detectable signal. The value thus obtained is normalized by the stellar luminosity $F_*$, which is estimated from the same dataset by means of the inner 1\arcsec \ of the total intensity $I$ image. The $I$ image is generated during the standard data reduction from the sum of the two beams with orthogonal polarization states \citep[see][]{Avenhaus2014a}. In summary, the disk-to-star contrast used for the analysis can be expressed as
\begin{equation} \label{Formula_albedo}
\phi_{\rm pol}= \frac{1}{r_{\rm out}-r_{\rm in}} \cdot \int_{r_{\rm in}}^{r_{\rm out}} \frac{F_{\rm pol}(r)\cdot 4\pi r^2}{F_*} dr 
.\end{equation}

This quantity, sometimes referred to as geometric albedo, is the combination of both the intrinsic albedo (affected by the specific dust properties) and the disk geometry (corresponding, along the disk major axis, to the disk flaring angle). The primary error on this estimate is computed from the $Q_{\phi}$ image by means of the weighted standard deviation on the resolution element around each datapoint, which is then propagated to $\phi_{\rm pol}$. To define an upper limit of non-detections, we carried out the same procedure with the cut on the $Q_{\phi}$ image being obtained from four averaged random directions. We found, nonetheless, that the error thus obtained is marginal compared to other systematic uncertainties, which we discuss here. 

First of all, the stellar halo used to compute $F_*$ also contains the (unseen) disk contribution. We consider this effect negligible since the brightest disks in our sample (with the exception of HD142527) contributes to roughly $1\%$ of the stellar brightness (assuming a conservative polarization fraction of $10\%$). In Herbig Ae/Be stars, a fraction of the near-IR flux may originate in the hot inner disk rather than in the stellar photosphere. However, part of this near-IR emission may also contribute to the illumination of the outer disk. Therefore, we do not correct for it by means of the photometric excess. Some of our intensity images have their inner few pixels saturated or covered by the coronagraph. In most of these cases, complementary frames with shorter integrations and without the coronagraph are available. Where these are not available, we estimate the missing inner photons by means of the dataset more similar in target, setup, and weather conditions. This procedure should not introduce an uncertainty larger than $10\%$. In the case of stellar companions visible from the image, we exclude the surrounding region and substitute it with the specular one.  

Secondly, a fraction of the scattered light from the disk may not be registered in the $Q_{\phi}$ images because of the deviations from azimuthal scattering to be expected from inclined disks. In fact, these may act to \textit{transfer} some signal from the $Q_{\phi}$ to the $U_{\phi}$ image. This may bias our estimates on inclined disks by an unpredictable but yet not dramatic fraction. Also, the smearing effect described by \citet{Avenhaus2014b} may damp the polarized flux of the inner $\sim 0.3\arcsec$ and result in a small underestimate of the contrast. 

Finally, and more importantly, the outermost radius with detectable signal from the PDI image cannot be defined univocally. Here we chose the location where the polarized flux drops below $3\sigma$, as calculated from the above-mentioned primary error on the image. However, we note that changing this radius by a fraction of an arcsecond results in a significant change to the computed contrast. The error bars listed in Table \ref{Table_sample}.1 and used through the paper take this uncertainty (which is by far the most significant) into account. On the other hand, the other uncertainties described above are not included and could account for an additional 20$\%$ systematic error on the contrast.

\end{appendix}

\end{document}